# DYNAMICAL STUDY OF A SECOND ORDER DPCM TRANSMISSION SYSTEM MODELED BY A PIECE-WISE LINEAR FUNCTION

Ina Taralova [1], Danièle Fournier-Prunaret [2]

[1] IRCCyN, UMR CNRS 6597, Ecole Centrale de Nantes, 1 rue de la Noë, B.P. 92101, 44321 Nantes Cedex 3, France.

[2] SYD-LESIA, DGEI, INSA, 135 avenue de Rangueil, 31077 Toulouse Cedex 4, France.

**Abstract: This paper analyses the behaviour of a second order DPCM (Differential Pulse Code Modulation) transmission system when the nonlinear characteristic of the quantizer is taken into consideration. In this way, qualitatively new properties of the DPCM system have been unravelled, which cannot be observed and explained if the nonlinearity of the quantizer is neglected. For the purposes of this study, a piece-wise linear nondifferentiable quantizer characteristic is considered. The resulting model of the DPCM is of the form of iteration equations (i.e. map), where the inverse iterate is not unique (i.e. noninvertible map). Therefore the mathematical theory of noninvertible maps is particularly suitable for this analysis, together with the more classic tools of Non Linear Dynamics. This study allowed us in addition to show from a theoretical point of view some new properties of nondifferentiable maps, in comparison with differentiable ones. After a short review of noninvertible maps, the presented methods and tools for noninvertible maps are applied to the DPCM system. An original algorithm for calculation of bifurcation curves for the DPCM map is proposed. Via the studies in the parameter and phase plane, different nonlinear phenomena such as the overlapping of bifurcation curves causing multistability, chaotic behaviour, or multiple basins with fractal boundary are pointed out. All observed phenomena show a very complex dynamical behaviour even in the constant input signal case, discussed here.**

## 1. INTRODUCTION

Differential Pulse Code Modulation (DPCM) transmission systems are widely used in telecommunications, speech and image coding [Dong H.K. et al, 1992], digital systems, medical research, signal processing [Bellanger, 1989], [Macchi & Uhl., 1993] and so on. Many papers have been devoted to this subject for a one step [Dinar, 1994], [Uhl & Fournier-Prunaret ,1995] or a two step [Fournier-Prunaret & al., 1993], [Gicquel, 1995] predictor. Different quantizer models have already been studied [Gicquel, 1995], [Fournier-Prunaret & al., 1993], [Uhl & al., 1991], but all of them were differentiable ones; the present paper claims to be a first attempt to model the quantizer characteristic by a piece-wise linear, i.e. a non differentiable, function. In this sense, the choice of the model puts this work into the field of digital filters with saturation-type overflow characteristic [Chua & Lin, 1988] [Ogorzalek, 1991], [Ogorzalek & Galias, 1991]. The first aim of this article is to compare our results (with the non-differentiable quantizer characteristic) to previous results (with the differentiable one), and further to compare these results with digital filters. Our second objective is to demonstrate the complexity of the system and the specific bifurcations taking place in the piece-wise linear case.

DPCM (see Fig.1) is a digital data compression technique based on error transmission. The encoder encodes through a quantization an input signal, which must be reconstructed at the output of the decoder. The differential part of the DPCM system is used to reduce the signal flow before its A/D transmission and is based on the following coding principle: as the successive signal values are usually correlated, it would be useful to make use of this transmission redundancy, without in any way losing information. Therefore, rather than transmitting the signal itself, only the prediction error $e_n$, i.e. the difference between the predicted and the effective signal values, $s_n$ and $\hat{s}_n$, is quantized and transmitted. The difference between the two correlated signals is thus coded in a smaller number of bits, and so the transmission data flow is reduced. The input signal $s_n$ has to be reconstructed at the end of the chain. The reconstructed signal is called $s'_n$. At the encoder, the predicted value $\hat{s}_n$ of the input signal $s_n$ is calculated based on its past samples by the recursive linear filter R. The encoder of the system (Fig.1) includes a nonlinear element - the quantizer. The latter is approximated by a piece-wise linear characteristic containing points of non-differentiability.

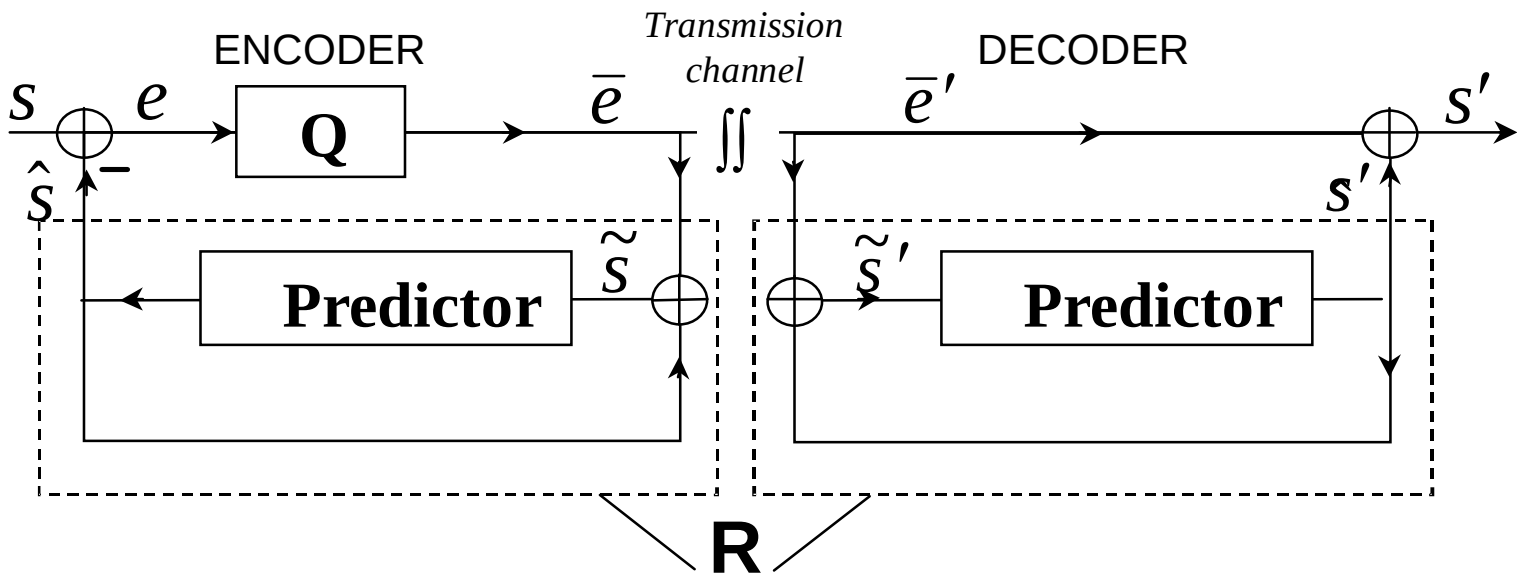

Fig. 1 : DPCM Transmission system

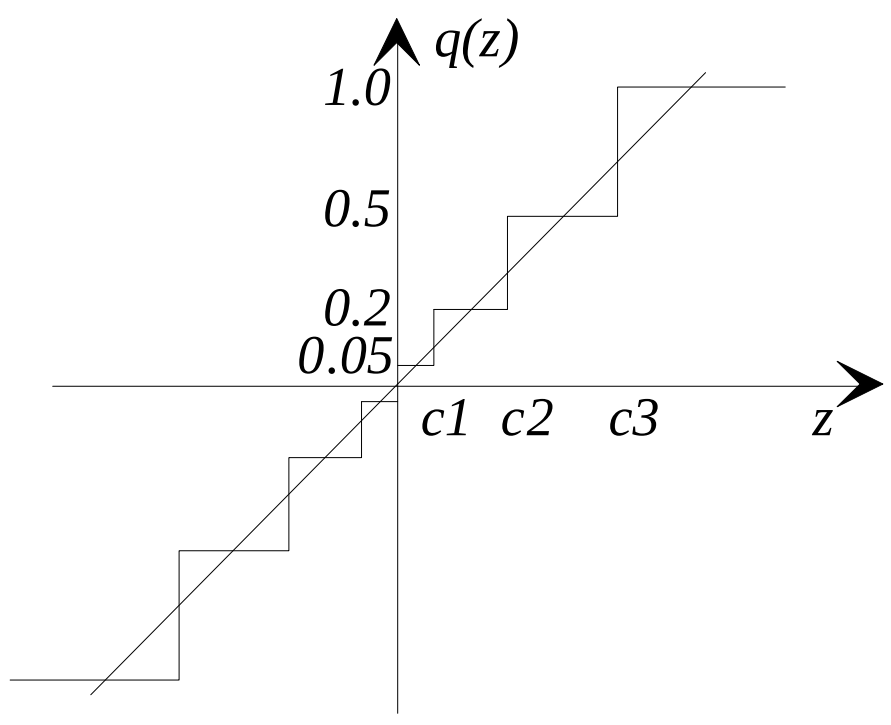

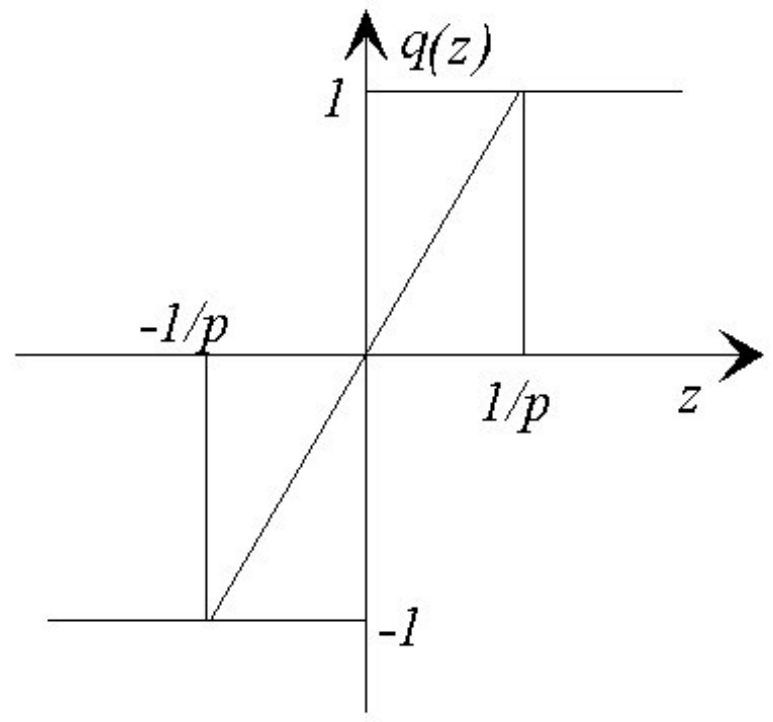

Fig.2a : Quantizer staircase characteristic ($c1$=0.1,$c2$=0.3 $c3$=0.7)

Fig.2b : Quantizer piece-wise linear characteristic p is the slope of the characteristic (p> 1)

The decoder is a linear system, therefore the nonlinear behaviour of a DPCM system is generated by the encoder and hereafter we shall focus on its dynamics. The quantizer of a DPCM encoder can be approximated as shown in Figs. 2a and 2b. Unlike previous studies, these models take into account the nondifferentiability of the quantizer characteristic and reveal new features of the driven system. In this paper, we have chosen to study the case of Fig. 2b; p is the slope of the quantizer characteristic, it is called companding gain and is larger than 1 (see [Uhl & al., 1991]). Despite the simple form of this piece-wise linear characteristic, we shall see that:

1) many new dynamical phenomena ensue from this model,
2) well-known phenomena, such as appearance of stable orbits, occur in a different way compared to the differentiable quantizer characteristic case.

Although more difficult for analytical studies, the characteristic of Fig.2a is closer to the real one and can be used for comparison. It has been shown in [Taralova-Roux & Fournier-Prunaret, 1996a] that the system behaviour is very similar in the cases of piece-wise linear and staircase characteristic.

This paper is organized as follows: Section 2 gives a short review of two-dimensional maps and the principal tools of noninvertible maps. Section 3 shows some basic properties of the DPCM system modelled by a noninvertible map. Section 4 focuses on typical nonlinear features exhibited by the DPCM system and analyses some typical cases. Section 5 summarizes the studies realized in the phase plane. To end the paper, section 6 concludes with final remarks and suggestions for future work.

# 2. OVERVIEW OF NONINVERTIBLE MAPS

## 2.1. CRITICAL LINES

Let the *map* $T: R^2 \to R^2$ be defined by the following system of equations :

$$\begin{cases} x_{n+1} = f(x_n, y_n) \\ y_{n+1} = g(x_n, y_n) \end{cases} \qquad (1)$$

where $x_n$, $y_n$ are real variables, $(x_n, y_n) \in \mathrm{R}^2$, and *f*, *g* are single valued piece-wise linear functions. In the following, we will denote the map *T PWL* for *piece-wise linear*. $(x_n, y_n)$ will also be denoted $X_n$.

In general we will say that the map $T$ is *invertible* if to any point $(x_{n+1}, y_{n+1})$ corresponds one and only one point $(x_n, y_n)$ which is called the *rank-1 preimage* of the point $(x_{n+1}, y_{n+1})$. Then the *inverse map* $T^{-1}$ has unique determination everywhere in the phase plane. The map $T$ is said to be *noninvertible* if there exist $(x_{n+1}, y_{n+1})$ with zero or more than one rank-1 preimages $(x_n, y_n)$.

Thus, when $T$ is a continuous noninvertible map, the phase plane can be divided into different regions $Z_i$, in which each point has the same number $i$ of preimages. The regions $Z_i$ are generally delimited by particular curves called *critical curves* $LC$ [Gumowski & Mira, 1980][Mira, 1987], along which at least two rank-1 preimages merge. In the differentiable case, critical curves $LC$ are obtained by cancelling the determinant of the Jacobian of $T$. The curve of merging rank-1 preimages of $LC$ is denoted $LC_{-1}$. Although for simplicity we consider here the two-dimensional case, the critical *curve* tool retains the same insight and can easily be extended to higher order (predictor accounting for several past samples) systems as a critical surface. In the PWL case, critical curves are defined in a different way, which is explained in the section 3.

## 2.2. SINGULARITIES IN A TWO-DIMENSIONAL MAP. FOLIATION OF THE PARAMETER PLANE

In this paragraph, we recall some results about the singularities of a two-dimensional map and their nature:

- A $k$-cycle (order $k$ cycle or period $k$ orbit) of $T$ consists of $k$ consecutive points (iterates or images) ($X_i$), $i=1,...,k$ satisfying $X_i = T^k X_i$ with $X_i \neq T^h X_i$, for $1 \le h < k$, $h$ and $k$ being integers. In other words, a cycle is a periodically repeating sequence of states $X_n$. A fixed point is a cycle with $k=1$. Figures 3a-3b show the representation of an order-3 cycle.
- Another index $j$ is also associated with a cycle in order to distinguish cycles having the same period $k$. This index characterizes the permutation of the $k$ cycle points by successive applications of the map $T$ [Mira, 1987].
- Let $X^*$ be a fixed point of $T$ and ($X_i$), $i=1,...,k$ be an order $k$ cycle, then their stability can be determined using the eigenvalues of the Jacobian matrix $DT(X^*)$ or $\prod_{i=1}^{k} DT(X_i)$. Let $S_l$, $l=1,2$, be the two eigenvalues, also called *multipliers*. Now, if the multipliers are real and :

  a) $|S_1| > 1$ and $|S_2| < 1$, $X^*$ or the $k$-cycle ($X_i$), $i$=1,...,$k$ is a saddle,

  b) $|S_1| > 1$ and $|S_2| > 1$, $X^*$ or the $k$-cycle ($X_i$), $i$=1,...,$k$ is an unstable node,

  c) $|S_1| < 1$ and $|S_2| < 1$, $X^*$ or the $k$-cycle ($X_i$), $i$=1,...,$k$ is a stable node.

If the multipliers are complex conjugates, i.e. $S_1=\rho e^{+i\varphi}$, $S_2=\rho e^{-i\varphi}$, then

a) $\rho > 1$, $X^*$ or the $k$-cycle ($X_i$), $i$=1,...,$k$ is an unstable focus

b) $\rho < 1$, $X^*$ or the $k$-cycle ($X_i$), $i$=1,...,$k$ is a stable focus

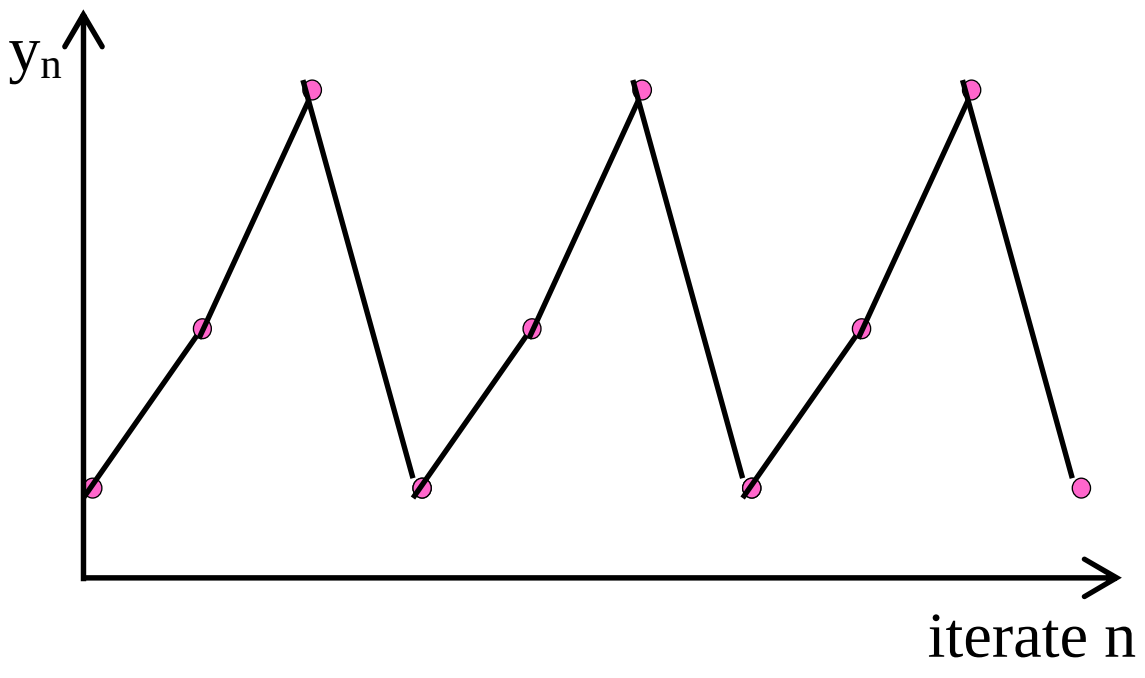

Fig. 3a : Order 3 periodic signal in the time domain.
periodic

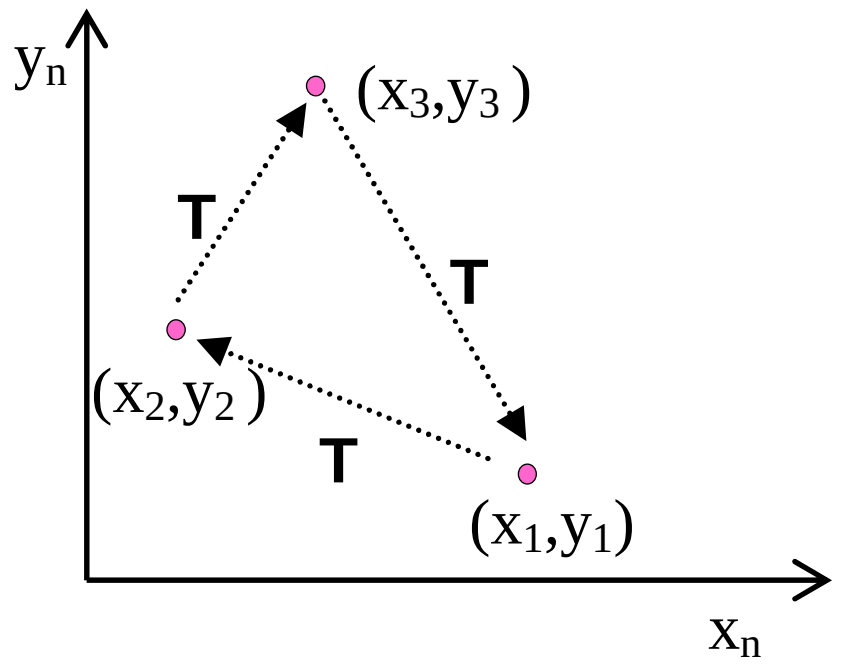

Fig. 3b : Corresponding order 3 orbit in the phase plane

A set of points, say *A*, is said to be invariant under *T* if it is exactly mapped into itself, that is,

$$T\,(A) = A \qquad (2)$$

We define an attractor A , for which there exists a non-zero measure set of initial conditions giving rise to iterated sequences converging towards A by application of the map *T*. Stable fixed points, stable period *k* orbits and asymptotically stable invariant closed curves are attractors. We also define a *chaotic attractor* from a practical point of view as bounded steady state behavior that is not an equilibrium point, not periodic state and not quasi-periodic. There is no widely accepted definition of chaos, from a mathematical point of view. A chaotic set may contain infinitely many unstable periodic orbits.

The set of initial conditions giving rise to iterated sequences converging towards a given attractor is called the *basin* of the attractor. A basin *B* is such that $T\,(B) \subset B$.

*T* can depend upon parameters; for instance, let us consider (*a*, *b*), two real parameters. When *a* and *b* are varied, one can observe qualitative change in the system. Such changes are called *bifurcations*. Let us define some particular bifurcations. The first one is called a *fold bifurcation*, which is denoted $\Lambda^{j}_{(k)_0}$ and corresponds to the appearance/disappearance of node and saddle periodic orbits according to the following scheme:

$$\begin{array}{l} \varnothing \longleftrightarrow \text{order } k \text{ stable node} + \text{order } k \text{ saddle} \\ \text{or} \\ \varnothing \longleftrightarrow \text{order } k \text{ unstable node} + \text{order } k \text{ saddle} \end{array} . \qquad (3)$$

where $\varnothing$ means absence of order *k* cycle.

In the differentiable case, a fold bifurcation curve is such that only one multiplier of an order *k* periodic orbit is $S = +1$, and corresponds in the simplest case to the merging of a cycle with one multiplier $S < 1$ with a cycle with one multiplier $S > 1$ . When a system is not differentiable, bifurcation curves have to be calculated in a non-traditional way : periodic orbits appear, following the scheme (3) when getting into contact with the lines of nondifferentiability (see section 4).

Similarly, a Neïmark-Sacker bifurcation curve $\Gamma_k^j$ is defined by $S_{1,2} = e^{\pm i\varphi}$ and corresponds to the generation of either one (simplest case) or several [Mira, 1969] [Mira, 1987] invariant closed curves (ICC) from a focus cycle. In the simplest case, the bifurcation scheme is:

$$\begin{array}{l} \text{order } k \text{ stable focus} \underset{\rho=1}{\longleftrightarrow} \text{order } k \text{ unstable focus} + \text{stable ICC} \\ \text{or} \\ \text{order } k \text{ unstable focus} \underset{\rho=1}{\longleftrightarrow} \text{order } k \text{ stable focus} + \text{unstable ICC} \end{array} \qquad (4)$$

Particular features of this bifurcation in the PWL case presented here are discussed further in section 4.

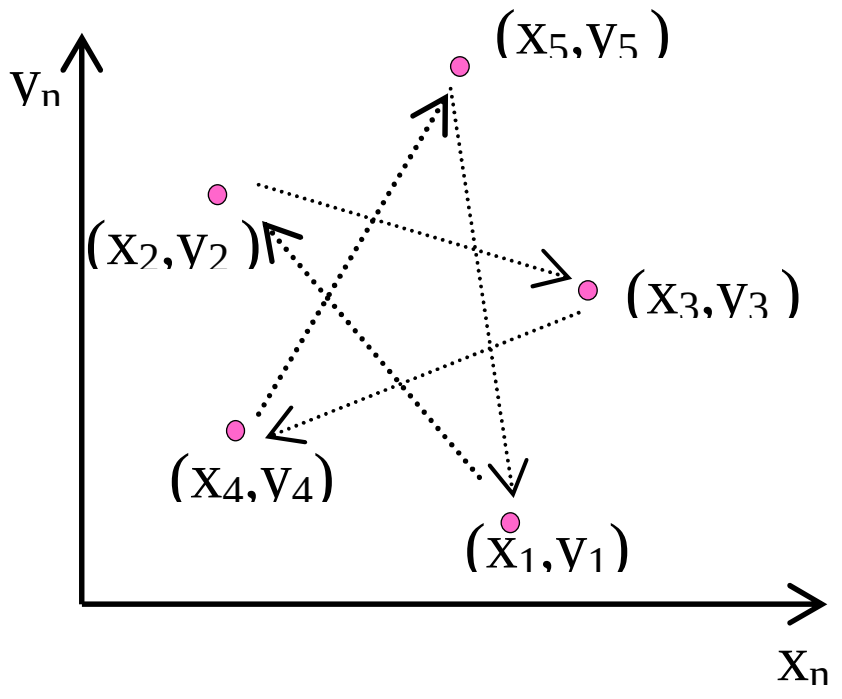

Fig. 3c : Period-5 cycle, rotation number 2/5.

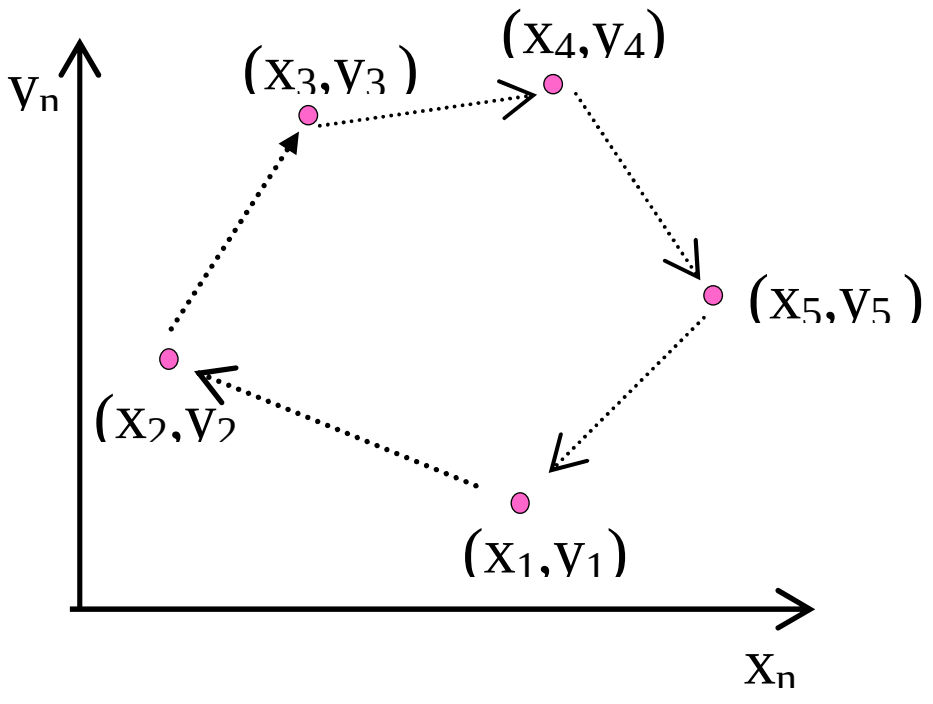

Fig. 3d : Period-5 cycle, rotation number 1/5.

Next, two cycles may have the same period *k*, without arising from the same bifurcation. A way to distinguish these cycles is to analyse the permutation mechanism of their *k* points, which is done by iterating consecutively *k* times in the phase plane, starting from any point of the cycle. When period-k cycles are issued from a Neïmark-Sacker bifurcation, one can define the associated rotation number *j/k*, which characterises a cycle in a unique way [Mira, 1987] (see Figs. 3c-d). In the rotation number *j/k*, the denominator *k* corresponds to the periodicity of the orbit, and the numerator *j* to the number of rotations necessary to pass through all points of the orbit; after *j* rotations the orbit comes to its initial point.

Each cycle exists in a precise domain of the parameter plane. Let us call such domain a *box*. The *box* is delimited by the *bifurcation curves*, between which the cycle exists. The bifurcation structure called *boxes in file* (for one-dimensional systems) is characterised by the typical ordering of the rotation numbers on the parameter axis, which exhibits a typical fractal structure, known also as the “devil’s staircase” structure. For two-dimensional systems, this ordering can be observed in the two-parameter plane, and the phenomenon is similar to Arnold tongues of the differentiable case because of the particular tongue-shaped form of the boxes [Mira,1987] [Boyland, 1986].

In the parameter plane, a given point is generally related to different cycles of order $k$ with possibly different multipliers and different values of rotation numbers. Then, this plane can be considered foliated and made up of sheets, each of them corresponding to a specific cycle. If $(x_n, y_n)$ is a point of such a cycle, and if we consider one of its coordinates, for instance $y_n$, then the $(a, b, y_n)$ space gives a qualitative three-dimensional representation of this sheet structure. In this foliated space, the sheets present folds along fold bifurcation curves. A fold curve joins two sheets, one related to the stable cycle born as a node, the other to the saddle one. The Fig. 3e gives an example of this situation with a fold bifurcation curve.

More complex communications between sheets, related to the existence of cusp points on bifurcation curves, also occur. For more details, see [Mira, 1987] [Mira & Carcassès, 1991] [Mira & al., 1991].

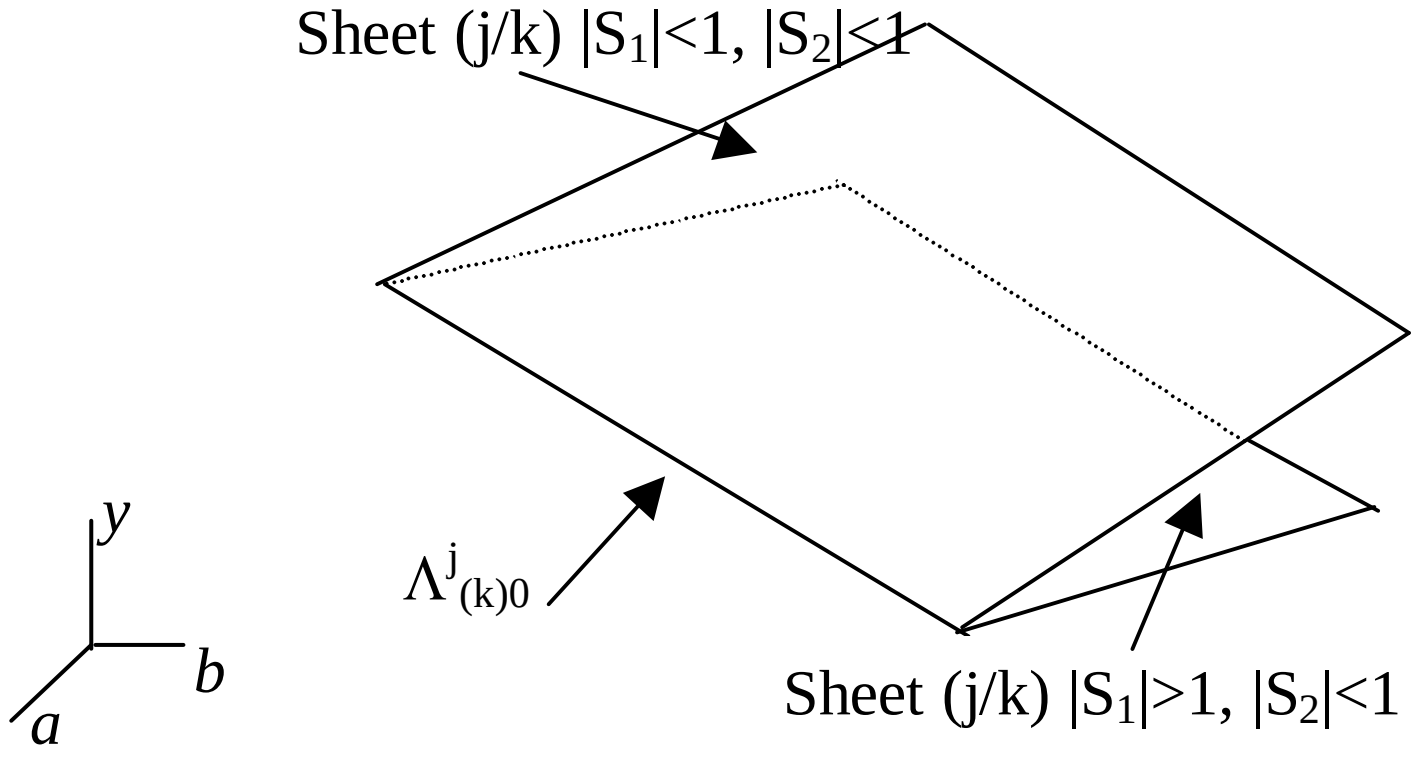

Fig. 3e : Disposition of the sheets of the phase-parameter plane in presence of a fold bifurcation curve.

## 3. SYSTEM MODELLING BY Z1/Z3/Z1 NONINVERTIBLE MAP AND CORRESPONDING PHASE SPACE FOLIATION

Let us consider now the DPCM system.

Considering what has been written in Section 1, the encoder input is the signal $s_n$ to be transmitted, the encoder output is the quantized version $\bar{e}_n$ of the prediction error $e_n = s_n - \hat{s}_n$, where $\hat{s}_n$ is the predicted signal. The signal $\tilde{s}_n = \hat{s}_n + \bar{e}_n$ is the reconstructed signal. For instance,

$$\hat{s}_{n+1} = a_1 \tilde{s}_n + a_2 \tilde{s}_{n-1} \qquad (5)$$

in the case of an order 2 predictor, $(a_1, a_2)^T$ being the predictor vector of parameters. The quantizer characteristic is taken as a piece-wise linear function (cf. Fig. 2b) :

$$q(s-z)=\begin{cases} p(s-z) & if\ \ |s-z|<\dfrac{1}{p} \\ sign(s-z) & if\ \ |s-z|>\dfrac{1}{p} \end{cases} \tag{6}$$

$p$ is called companding gain and is always larger than 1, it serves to increase the signal/noise ratio; it corresponds to the slope of the piece-wise linear quantizer approximation presented in Fig. 2b.

In previous studies, the quantizer has been modeled by the following differentiable characteristic [Macchi & Uhl, 1993], [Fournier-Prunaret & al., 1993] :

$$q(s-z)=\tanh p(s-z) \tag{7}$$

The decoder predictor has the same parameter vector $(a_1, a_2)^T$ as the encoder. In the absence of transmission errors at time « $j$ » $(\bar{e}'_j=\bar{e}_j, j\le n)$, if the decoder is suitably initialized $(\tilde{s}'_0=\tilde{s}_0, \tilde{s}'_1=\tilde{s}_1)$ its output $\tilde{s}'_n$ at time « $n$ » is the signal $\tilde{s}_n$ reconstructed by the encoder. Thus

$$s_n-\tilde{s}_n=e_n-\bar{e}_n \tag{8}$$

which means that the transmission noise $s_n-\tilde{s}_n'$ is merely the quantization noise. Here the order 2 predictor is investigated, in the particular case of a constant input signal $s_n = s$. This assumption is often used in the literature and can be justified by the fact that at high sampling rates a time-varying input can be approximated by a dc input over relatively large time intervals. In fact, the assumption of $s$ not constant leads to a higher order autonomous system [Rouabhi & Fournier-Prunaret, 1999], whose analysis would be highly intricate.

As the decoder is a linear system, we will concentrate from this point on the analysis of the DPCM encoder. One can write :

$$\hat{s}_{n+1}=a_1\hat{s}_n+a_1q(s-\hat{s}_n)+a_2\hat{s}_{n-1}+a_2q(s-\hat{s}_{n-1}) \tag{9}$$

A corresponding substitution can be applied to the DPCM encoder by selecting new variables: $x_n=\hat{s}_{n-1}$ and $y_n=\hat{s}_n$, thus $x_n$ et $y_n$ correspond to two consecutive estimations of the input signal s. This substitution transforms the second order difference equation (9) into two first order recurrent equations, giving rise to the following two-dimensional noninvertible nondifferentiable map:

$$X_{n+1}=T(X_n) \quad \text{where} \quad X_n=\begin{pmatrix} x_n \\ y_n \end{pmatrix} \tag{10}$$

$$T:\begin{cases} x_{n+1} = f(x_n, y_n) = y_n \\ y_{n+1} = g(x_n, y_n) = a_2\big(x_n + q(s - x_n)\big) + a_1\big(y_n + q(s - y_n)\big) \end{cases} \qquad (11)$$

$q$ satisfying equation (6).

As mentioned at the beginning of this section, the state variables $x_n$ and $y_n$ correspond to two consecutive estimations of the input signal *s* which is constant and so considered to be a parameter for this study. In this case the function *g* is no longer continuously differentiable, and particular properties appear that do not arise in the continuously differentiable case [Macchi & Uhl, 1993], [Fournier-Prunaret & al., 1993].

In the next sections we will often refer to the critical lines of the DPCM, so let us calculate them first. There is a difference in principle between the differentiable and the non-differentiable case. It is not possible to apply the classical definition for critical lines determination implying Jacobian determinant canceling, as the latter is not defined at the lines along which the determination of the piece-wise linear map changes. Moreover, the eigenvalues change with discontinuity when crossing through these lines. For this reason, another method described in [Mira & Gumowski, 1966] has been used to define critical lines. This method is the following. In the phase plane $(x_n, y_n)$ two families of curves, corresponding to the right hand side of (11) are plotted: $x_{n+1} = f(x_n, y_n) = \alpha = const$ and $y_{n+1} = g(x_n, y_n) = \beta = const$, where $\alpha$ and $\beta$ are variable parameters (Fig.4). In this way, the phase plane represents the DPCM map both at time « *n* » and « *n+1*» : the axes $(x_n, y_n)$ give the present states of the system, i.e. at time « *n* », and the curves $\alpha$ and $\beta$ represent the next states of the system, i.e. at time « *n+1* ». Since for our model $x_{n+1} = y_n$, the first family of curves $x_{n+1} = f(x_n, y_n) = \alpha = const$ is in fact an infinity of parallel horizontal lines in the plane $(x_n, y_n)$ for all $y_n = const$ (these lines are not plotted in order to simplify the drawing). We can now determine the image $(x_{ni+1}, y_{ni+1})$ of the point $(x_{ni}, y_{ni})$ using the drawing: indeed, according to (11), this image (or iterate) is given by the right hand side of (11), i.e. $(x_{ni+1}, y_{ni+1}) = (f(x_{ni}, y_{ni}), g(x_{ni}, y_{ni})) = (\alpha_i, \beta_i)$. In the phase plane $(x_n, y_n)$, the curve $f(x_{ni}, y_{ni}) = \alpha_i$ intersects the curve $g(x_{ni}, y_{ni}) = \beta_i$ at the point with coordinates $(x_{ni}, y_{ni})$. Fig. 4 shows an example: let the image (or iterate) of the point $\mathrm{M}_1 = (x_{ni}, y_{ni})$ be the point $\mathrm{P} = (x_{ni+1}, y_{ni+1}) = T(\mathrm{M}_1)$; suppose we know $(x_{ni}, y_{ni})$ and we are looking for $(x_{ni+1}, y_{ni+1})$. If we read the values of the curves $\alpha_i$ and $\beta_i$ passing through $\mathrm{M}_1 = (x_{ni}, y_{ni})$, we obtain the coordinates of the image $\mathrm{P} = T(\mathrm{M}_1)$ shown in Fig.4 since $x_{ni+1} = y_{ni} = \alpha_i$ and $y_{ni+1} = \beta_i$. In general (i.e. when P is not a fixed point), the coordinates of P and $M_1$ are different. Now, suppose we are interested in the opposite problem, i.e. the determination of the inverses of P, knowing its coordinates $(x_{ni+1}, y_{ni+1})$. First, we have to localise in the plane $(x_n, y_n)$ the curves $\alpha_i$ and $\beta_i$ such as $\alpha_i = x_{ni+1} = y_{ni}$ and $\beta_i = y_{ni+1}$, then find their intersection(s), and finally project the intersection point(s) on the axes $(x_n, y_n)$. The resulting coordinates correspond to the preimage(s).

For our example, if we follow the curve $\beta_i$ on which the point $M_1$ is located, we can notice that besides the point $M_1$, there are two more intersection points between the curve $f(x_{ni}, y_{ni}) = \alpha_i$ and the curve $g(x_{ni}, y_{ni}) = \beta_i$, let us denote them by $M_2$ and $M_3$. All three distinct points $M_1$, $M_2$ and $M_3$ have the same image $T(M_1) = T(M_2) = T(M_3) = P = (x_{ni+1}, y_{ni+1}) = (\alpha_i, \beta_i)$, but correspond to different preimages: $M_1(x_{ni}, y_{ni})$, $M_2(x'_{ni}, y_{ni})$ and $M_3(x''_{ni}, y_{ni})$; thus $T^{-1}(P) = \{M_1, M_2, M_3\}$. We can see that the DPCM map is noninvertible, since to one couple of coordinates $(x_{ni+1}, y_{ni+1})$ correspond three different preimages. Now it is useful to find the location in the plane $(x_n, y_n)$ of the regions $Z_3$ with three preimages. According to the definition, the number of preimages changes when the critical lines are crossed, so we need to localise the critical lines. To do this, let us find the location of points where the number of intersections of α and β changes. If we consider again Fig. 4, for a given $\beta_i$ curve and because of the shape of the α and β curves, we observe either 1, or 3 intersections with a given curve $\alpha_i$. There always exists at least one preimage, i.e. at least one intersection point of $\alpha_i$ and $\beta_i$, since the lines $\alpha = const$ are parallel horizontal lines and they exist for any α. Now, if we imagine « moving » the curve $\beta_i$ up and down, the number of intersections of $\alpha_i$ and $\beta_i$ will change from 1 to 3 and vice versa, and this will happen along the lines :

$$|s - x_n| = \frac{1}{p} \qquad \text{(11a)}$$

where the determination of the map changes ; at the limit when $\alpha_i$ and $\beta_i$ are tangent, we have one separate preimage and two merging preimages (corresponding to the peak of the curve $\beta_i$). The lines with merging preimages are also the lines (11a). As mentioned in section 2, critical curves $LC$ are such that *along them* at least two preimages by point merge ; therefore, $LC$ can be constructed point by point, taking the images of the points where the curves α and β are tangential. Following this rule, two critical lines $LC_a$ and $LC_b$ have been constructed, as images of the two lines of discontinuity (11a) . The obtained critical lines separate the phase plane into regions with qualitatively different behaviour. Each point chosen between $LC_a$ and $LC_b$ possesses three rank-one preimages, and only one outside. The equivalent graphical representation is provided in the state plane, containing nine different regions $D_1$, $D_2$, … $D_9$, according to the lines of discontinuity of the derivatives (11a) and :

$$\left|s - y_n\right| = \frac{1}{p} \qquad (11b)$$

The lines (11b) are not associated with a difference in the number of preimages (and thus are not critical lines) while for the other two (11a), their images separate zones whose points have 1 or 3 rank-1 preimages. Thus these are critical curves $LC_{-1}$ and their images are $LC$ curves. The obtained critical curves $LC$ separate the phase plane into regions with qualitatively different behaviour. For the map (11), the critical curves are given by the following equations :

$$\begin{aligned} &\mathrm{LC}_{-1a} : \mathrm{x} = -\frac{1}{\mathrm{p}} + \mathrm{s} \\ &\mathrm{LC}_{-1b} : \mathrm{x} = \frac{1}{\mathrm{p}} + \mathrm{s} \end{aligned} \qquad (12)$$

$$\begin{aligned} &LC_a \begin{cases} y = a_1\left(ps + (1-p)x\right) + a_2\left(\frac{1}{p} + s - 1\right) & \text{if } \left|s - x\right| < \frac{1}{p} \\ y = a_1\left(x + sign(s-x)\right) + a_2\left(\frac{1}{p} + s - 1\right) & \text{if } \left|s - x\right| > \frac{1}{p} \end{cases} \\ &LC_b \begin{cases} y = a_1\left(ps + (1-p)x\right) + a_2\left(-\frac{1}{p} + s + 1\right) & \text{if } \left|s - x\right| < \frac{1}{p} \\ y = a_1\left(x + sign(s-x)\right) + a_2\left(-\frac{1}{p} + s + 1\right) & \text{if } \left|s - x\right| > \frac{1}{p} \end{cases} \end{aligned} \qquad (13)$$

We observe three regions in the $(x, y)$ plane with two different kinds of behaviour :

- ⇨ $Z_3$ between $LC_a$ and $LC_b$, where a point possesses three rank-1 preimages
- ⇨ two regions $Z_1$ outside, where a point possesses one rank-1 preimage

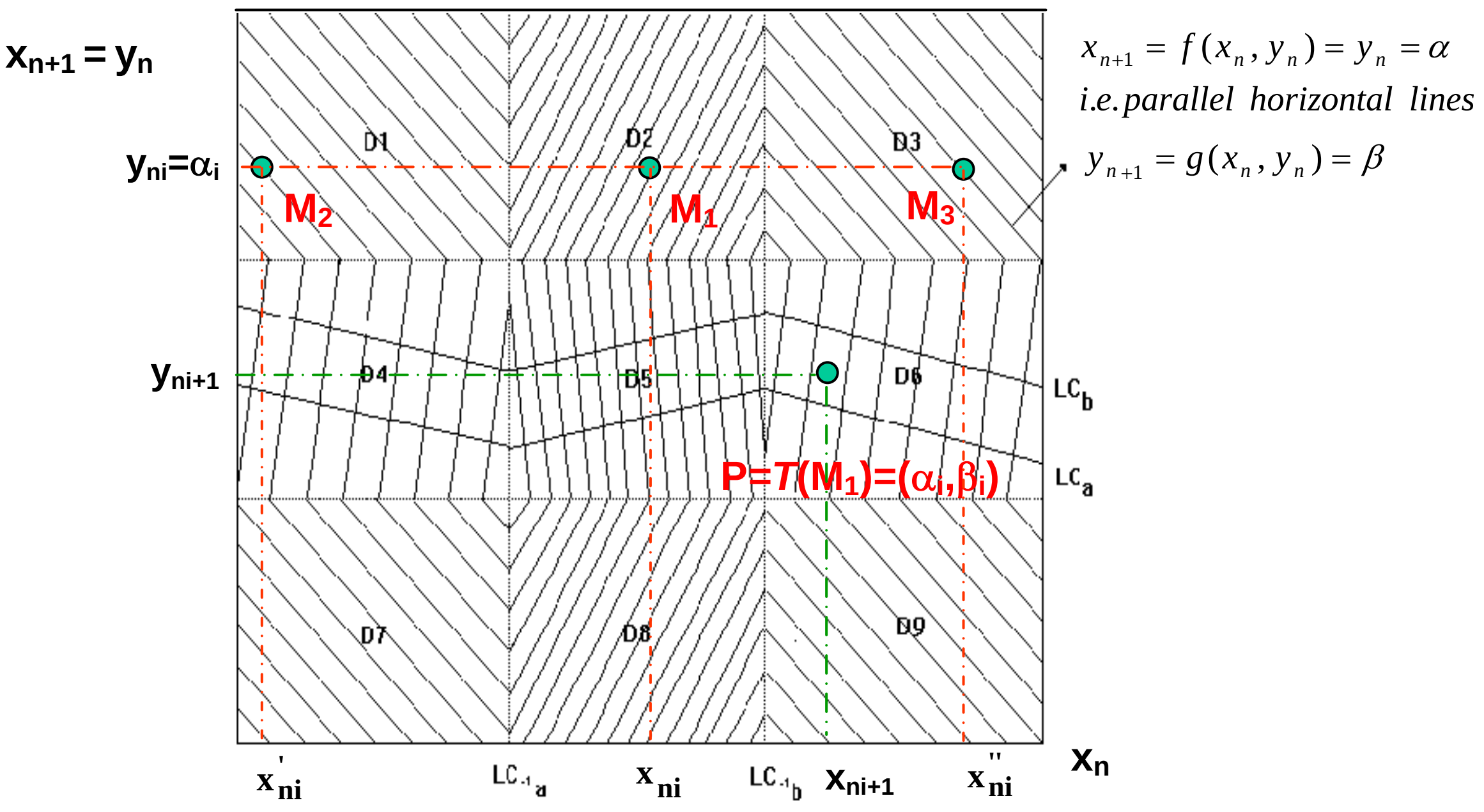

Fig. 4 : Numerico-analytical approach for calculating the critical lines

The equivalent graphical representation is provided in the phase plane, containing nine different regions according to the lines (11a) (11b) as shown in Fig. 4. These regions are associated with the different determinations (i.e. defining functions) of the map $T$ (11). According to (6) where z is either $x_n$ or $y_n$, there are three regions with different deteminations for $x_n$ and three regions with different deteminations for $y_n$ which correspond to nine regions D1,…,D9 with different determinations of the map $T$ in the plane $(x_n, y_n)$.

The *preimages* can be calculated as follows:

$$T_1^{-1}:\begin{cases} x_n = \{[y_{n+1} - a_1[x_{n+1} + q(s - x_{n+1})]\}/a_2 + 1 \\ y_n = x_{n+1} \end{cases} \qquad (14)$$

this inverse gives a point with $x_n > s + \frac{1}{p}$ i.e. $x_n \in D_3 \cup D_6 \cup D_9$ (14a)

$$T_2^{-1}:\begin{cases} x_n = \{[y_{n+1} - a_1[x_{n+1} + q(s - x_{n+1})]\}/a_2 - 1 \\ y_n = x_{n+1} \end{cases} \qquad (15)$$

this inverse gives a point with $x_n < s - \frac{1}{p}$ i.e. $x_n \in D_1 \cup D_4 \cup D_7$ (15a)

$$T_3^{-1}:\begin{cases} x_n = \{y_{n+1} - a_1[x_{n+1} + q(s - x_{n+1})] - a_2 ps\}/[a_2(1-p)] \\ y_n = x_{n+1} \end{cases} \qquad (16)$$

this inverse gives a point with $s - \frac{1}{p} < x_n < s + \frac{1}{p}$ i.e. $x_n \in D_2 \cup D_5 \cup D_8$ (16a)

These expressions have been obtained taking into account that $q(s - x_{n+1}) = q(s - y_n)$. For points $(x_{n+1}, y_{n+1})$ belonging to the strip between the two critical curves $LC_a$ and $LC_b$, which is the region $Z_3$, three inverses having the same $y_n$ exist $(x_{n1}, y_n), (x_{n2}, y_n), (x_{n3}, y_n)$, while for points above $LC_b$ and below $LC_a$, which is the region $Z_1$, only one inverse exists.

Fig. 4 also gives the two dimensional non-linear projection of a three-dimensional qualitative representation [Taralova-Roux & Fournier-Prunaret, 1998]. We are used to the two-dimensional phase plane representation. Nevertheless, if we calculate and draw the preimage $x_{n-1}$ as a function of $(x_n, y_n)$ it is possible to reconstruct a kind of three dimensional *foliation* of the phase space of the map (11) as shown on Fig. 5a and Fig. 5b according to the sign of the parameter $a_1$. The upper pictures show the phase plane, and the lower ones the corresponding three-dimensional foliation, with the representation of a point in $Z_3$ and of a point in $Z_1$. The foliation shows that a point in $Z_3$ is the image of three points and the point in $Z_1$ of one point. This representation enables a visual understanding of the notion of *noninvertibility* and better explanation of the phenomena arising in the *phase* plane.

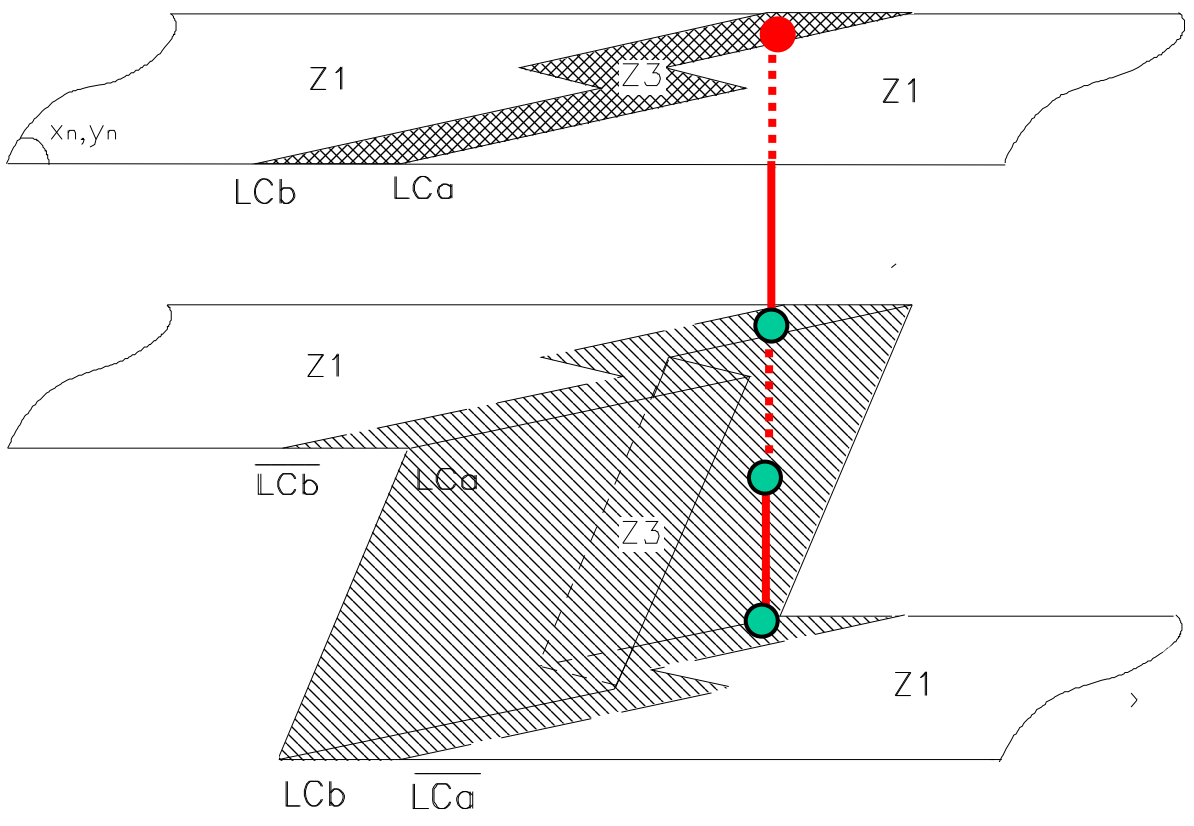

Fig.5a : Phase Plane Foliation for $a_1<0$. One can see that the red point located in the region $Z_3$ is the image of three distinct points (in green) while each point in the region $Z_1$ is the image of only one point.

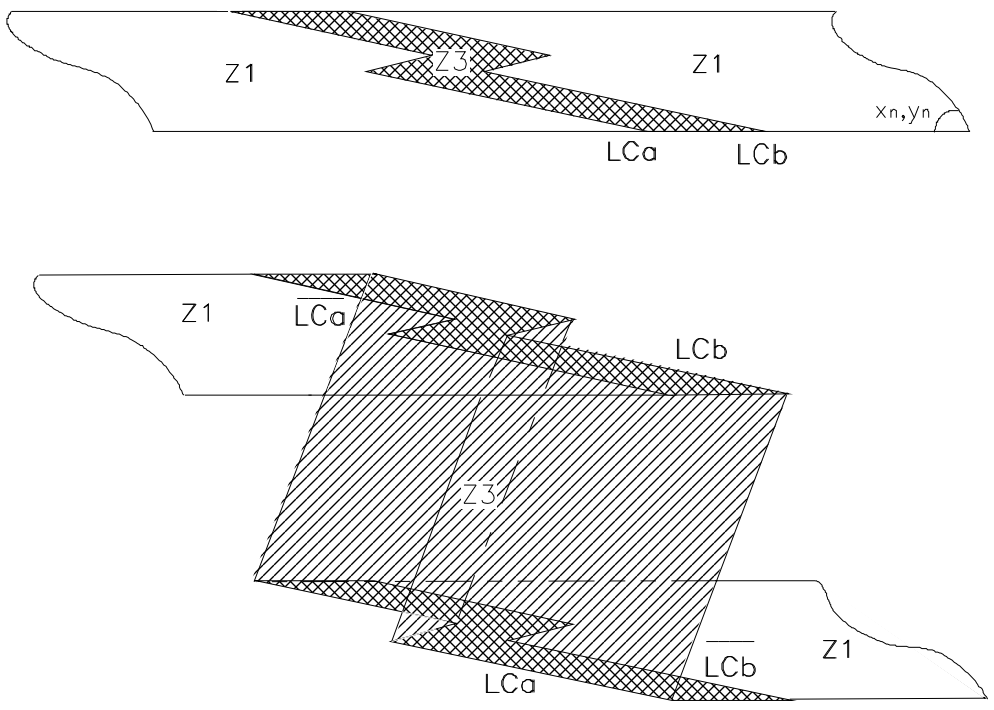

Fig.5b Phase Plane Foliation for $a_1>0$

Now, to show one of the uses of the critical lines tool, let us consider an attractor and its basin D, the boundary of which is crossed by $LC_a$ (Figs.6a, 6b). Let us call $H_0$ the part limited by segments of $LC_a$ and the boundary of the domain D. $H_0$ is inside the zone $Z_3$, two preimages of $H_0$ are located on the two sides of $LC_{a-1}$ forming a hole $H_1^{1,2}$ inside D. The third preimage $H_1^3$ is located in a region Z3; therefore $H_1^3$ possesses its own three rank-1 preimages $H_2^1$ , $H_2^2$ , $H_2^3$, which are spread in the phase plane, which have their rank-1 preimages etc . The resulting cheese-like domain can possess an infinite countable number of holes and is called *multiply connected*. Let us consider now two different basins $D_1$ (in pink) and $D_2$ (in red) which coexist (fig.7a, 7b). The critical line $LC_a$ "cuts" a piece of $D_1$, forming a domain $H_0$ and, following the properties of critical curves already used in the previous example, the preimage of $H_0$, denoted $H_1^{1,2}$, is found back on both sides of $LC_{a-1}$ forming respectively an *island* for $D_1$, and a *hole* for $D_2$. All original points ($x_n$,$y_n$) located inside as well as their rank "k" preimages (e.g. all rank "k" holes) do not belong to $D_2$, but to $D_1$. $D_2$ becomes a *multiply connected* domain and $D_1$ a *non-connected* domain. [Mira & al., 1996 ]

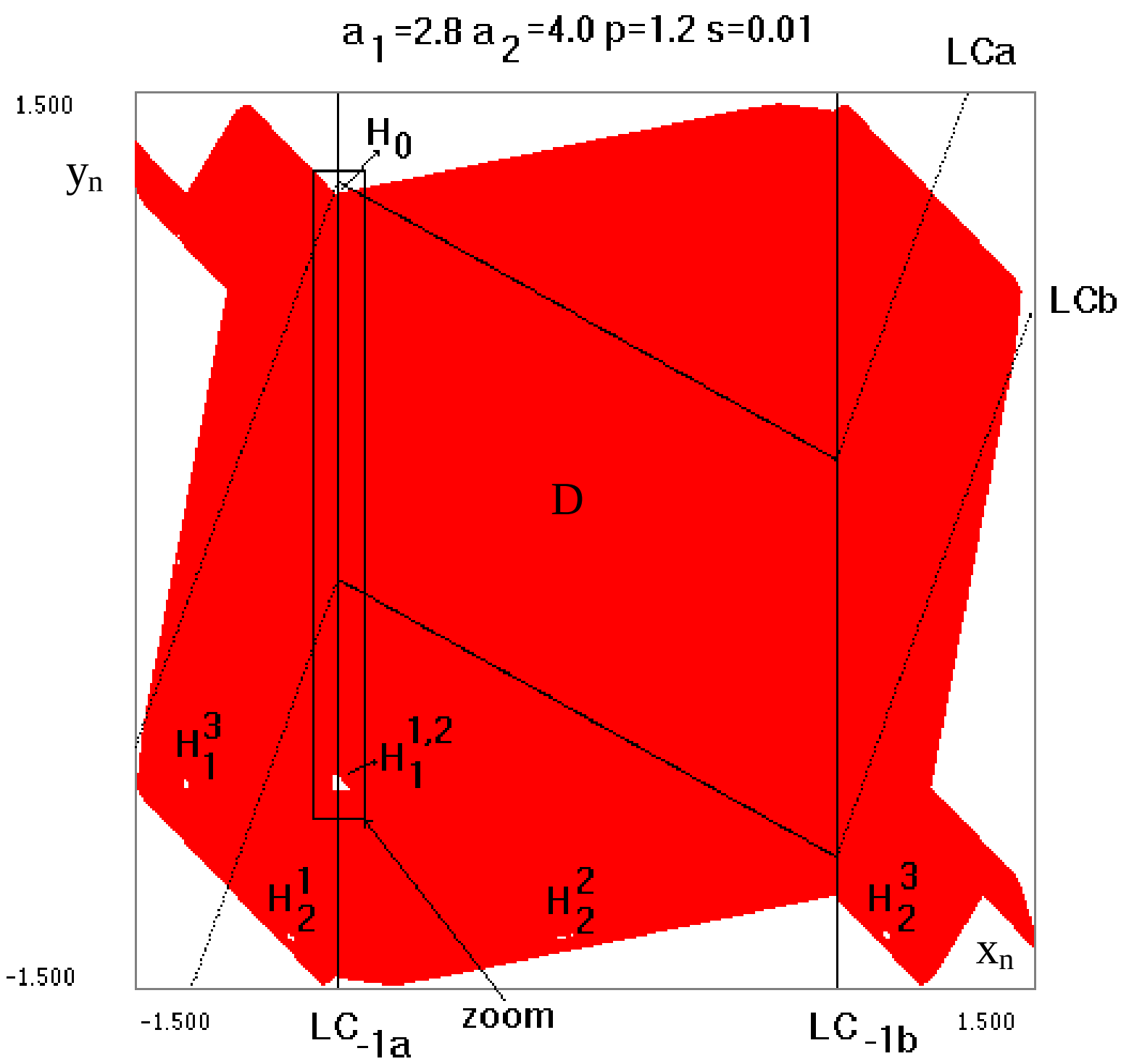

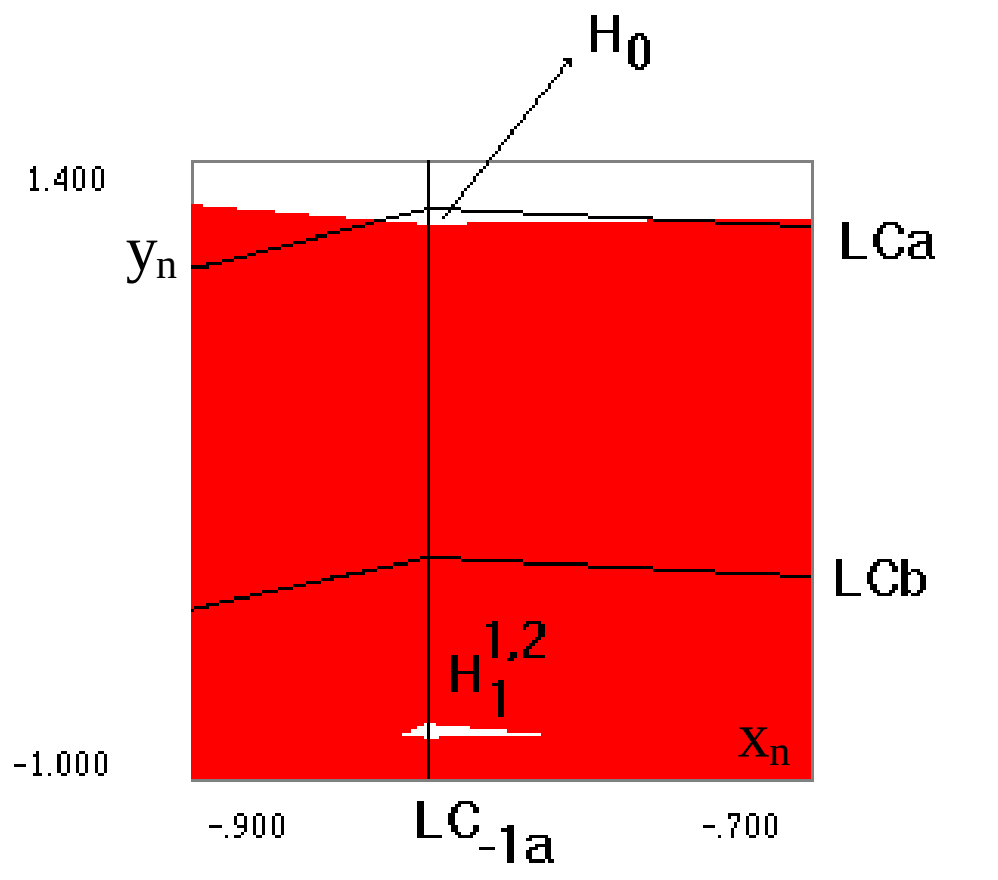

Fig.6a : Bifurcation connected → multiply connected basin

Fig.6b : Enlargment of Fig.6a

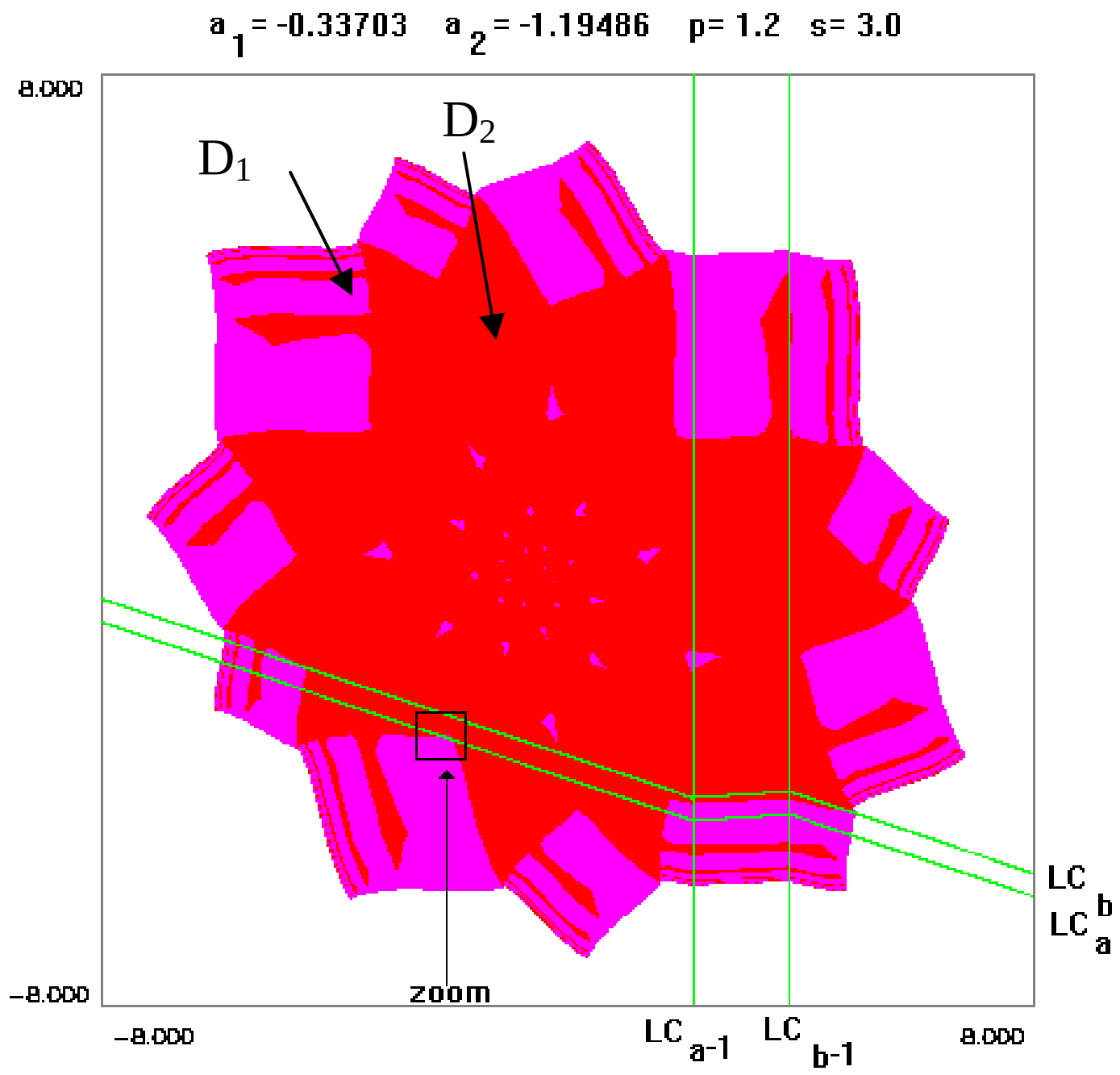

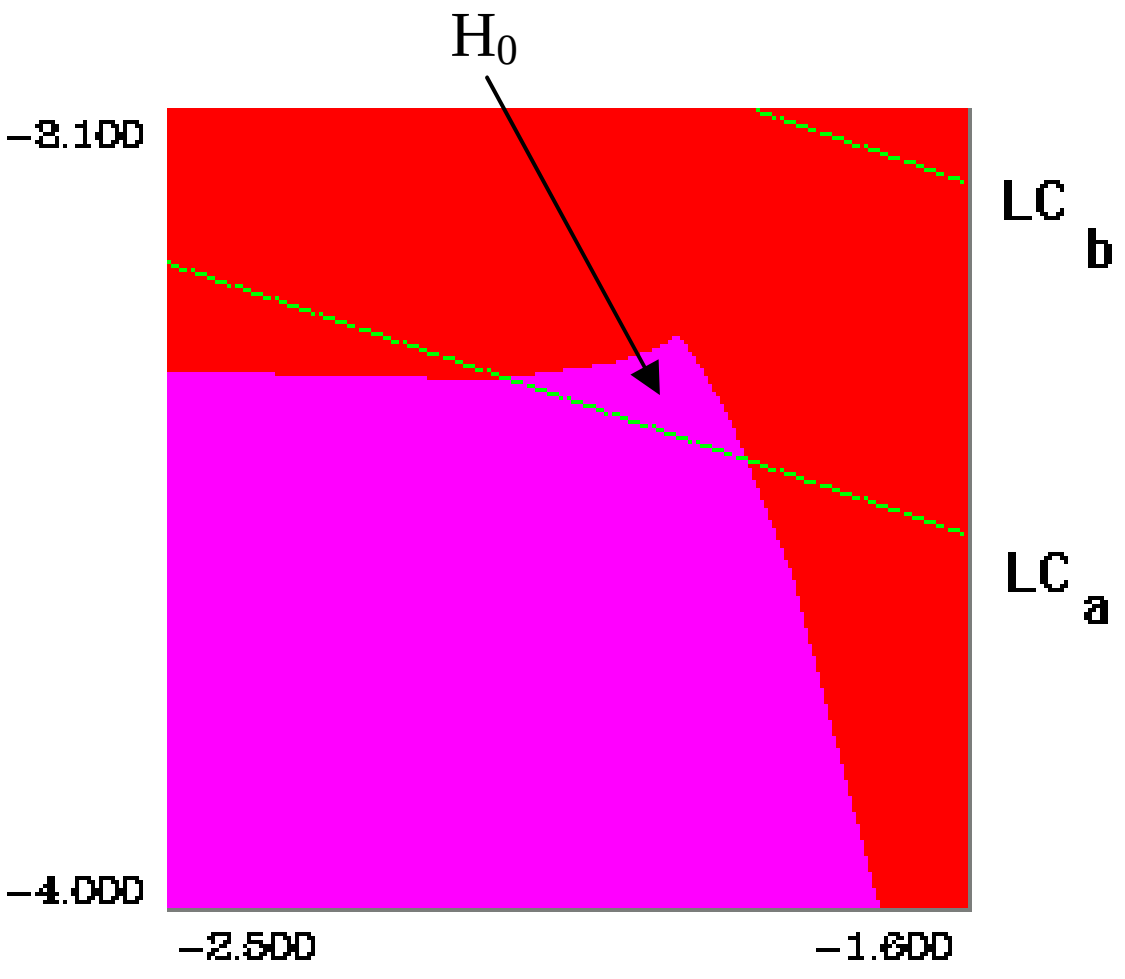

Fig.7a : In the phase plane ($x_n$, $y_n$), passage connected → multiply connected basin in the case where two different attractors coexist. Their basins are shown in red and pink.

Fig.7b : Enlargment of Fig.7a.

## 4. BIFURCATIONS

By bifurcation we denote qualitative change in the system behaviour; for instance, switching from one stable to another different stable state, from stable to unstable state, change in the geometrical shape of a basin, etc. Bifurcations occur when the system parameters are varied. They can be tracked by analyzing the influence of the variations in the parameter vector for a map $\boldsymbol{T}$ on its solutions.

The study of bifurcations is extremely important, especially when the model describes a transmission system and when its stability is concerned. Due to the tolerances and time-variant features of electronic components, it is well known that two physical circuits can never be identical; bifurcation studies give us the range of parameters and the conditions under which the behaviour of the circuit remains unaltered. An additional restriction in the case of the DPCM system is that the encoder and the decoder have to be lined up. The equality of the encoder input and the reconstructed decoder output signals is formally required. A necessary condition is the uniform encoder stability in time, independently of the adaptation type. The evolution of the signal depends on certain parameters. We seek to analyze this dependence and to identify significant changes that certain evolutions undergo as the parameters are varied. Variation in one or several components of the parameter vector may give rise to periodic orbits (cycles) at the quantizer output, or more complex phenomena (e.g. chaotic behaviour), structure modifications in the boundaries of basins in the phase plane, etc. Parameter vector values corresponding to these qualitative modifications are called *bifurcation values*. A study of these bifurcations forms the bulk of the present section.

## 4.1. BIFURCATION DIAGRAMS

When the parameter vector dimension is greater than one, bifurcations occur over a curve (surface in 3D) called a *bifurcation curve*. A bifurcation curve gives the region of admissibility of a given stable state. Bifurcation curves are summarized in *bifurcation diagrams* and tell us a priori for which parameter values the system converges or diverges, exhibits chaotic or periodic behaviour. Crossing a bifurcation curve implies qualitative change in the system behaviour.

All the bifurcation diagrams shown in Figs. 8-12 and Fig. 14 concern the DPCM model (map(11)); they are two-dimensional and they refer to the map (11) with different values of the companding gain $p$ and the input signal $s$ (constant for this study). As the parameter vector $[a_1, a_2, p, s]$ is four-dimensional, only the plane of predictor parameters $(a_1, a_2)$ has been scanned, with $p$ and $s$ values chosen according to their physical meaning: $p$ higher than one, and input signal amplitude $s$ involving full use of the quantizer characteristic (both linear part and saturation).

Stable solutions are obtained by iterating from different initial conditions $(x_0,\ y_0)$ for fixed parameter values $[a_1, a_2, p, s]$ . Each color corresponds to a region of existence of different stable behaviour (attractors): fixed point (blue), stable period-two orbit (green), stable period-three orbit (cyan) etc. The white area corresponds to divergent iterated sequences in the phase plane. The black area corresponds to $l$-periodic orbits ($l > 14$) or bounded iterated sequences including chaotic behaviour. Each different colour defines a region of admissibility of a $k$-periodic orbit ($k < 14$), delimited by a bifurcation curve. Any point $(a_1, a_2)$, chosen inside a given region and corresponding to the existence of a given periodic orbit, guarantees at least one solution of the considered order; nevertheless its structural stability, i.e. its robustness with respect to small $(a_1, a_2)$ perturbations must be studied more precisely via a more detailed analysis of the phase plane. In order to compare separately the influence of $p$ and *s*, the $(a_1, a_2)$ parameter plane has been scanned for fixed *p* and different *s* values (Figs. 8;9;12 and Figs. 10;11), and for fixed *s* and different *p* values (Figs. 8;10).

This representation gives a global view of the parameter plane. Figures 8-12 show areas in the $(a_1, a_2)$ plane (coloured parts of Figs. 8-12), where at least one attractor exists in the phase plane. On Figs. 8-12, a special domain, the stability triangle (τ) is represented. This triangle corresponds to the predictor parameters $(a_1, a_2)$ for which the DPCM transmission system is considered as stable and well-functioning by users. These results have been obtained using linear studies, by neglecting the nonlinearity of the quantizer. For our study, we have analyzed the DPCM encoder inside and outside the stability triangle. The latter study is justified, since setting the parameters outside the stability triangle may be used to break undesirable orbits; moreover, if an adaptive filter is used, during the adaptation algorithm the parameters can leave the stability triangle.

One can observe that, when *p* is small enough (*p*=1.2), the stability triangle only contains parameter values for which periodic orbits exist and, when p is higher (*p*=5), (τ) contains black regions with possible chaotic behaviour, which can disturb the functioning of the system [Macchi & Uhl, 1993] [Fournier-Prunaret & *al*., 1993].

One can see that the stability triangle is bounded on its lower part by the bifurcation curve $\Gamma_1^1$ (bifurcation (4) for a fixed point, see section 4.2).

From Figs.8-12, it can be seen that, as in the differentiable case [Fournier-Prunaret & *al*., 1993], when *p* is increasing, the parameter plane domain where an attractor exists has a decreasing size.

These diagrams are not exhaustive since several attractors may coexist for the same parameter values, as was shown in the previous section, due to the foliation of the parameter plane. Indeed, bifurcation curves may overlap. The mere knowledge of the existence of multistability is not sufficient to give information about the stability of a system since the coexisting stable solutions can be so far away from the set point in the phase plane to be unimportant in practice. We must quantify the stability of the various attractors to finite-amplitude perturbations by constructing their basins in the phase plane, that is to say the set of all points that asymptotically approach the attractor under forward iteration. This problem will be thoroughly analyzed in section 5.

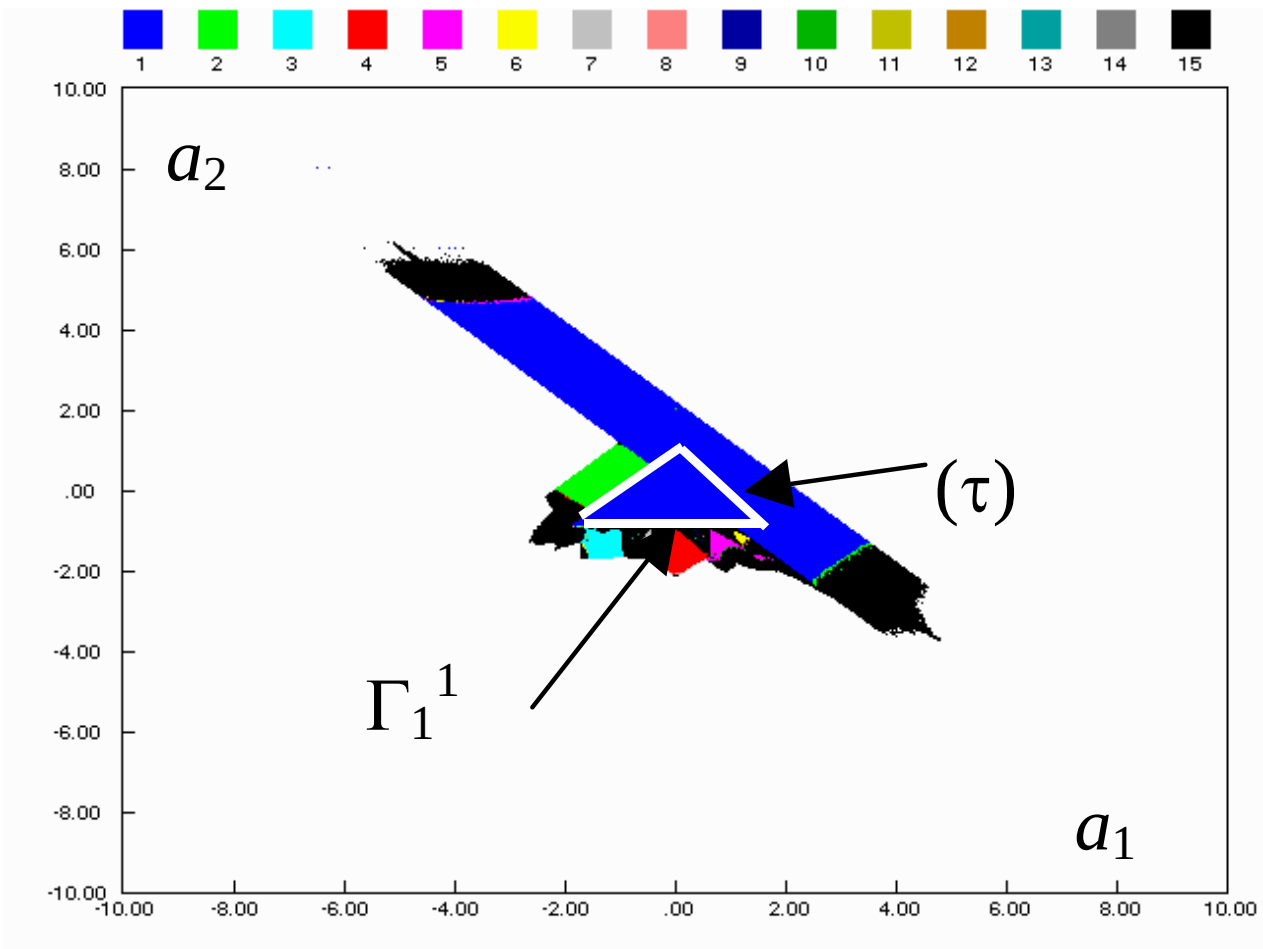

Fig. 8 : Parameter plane (p=1.2 s=0.5)

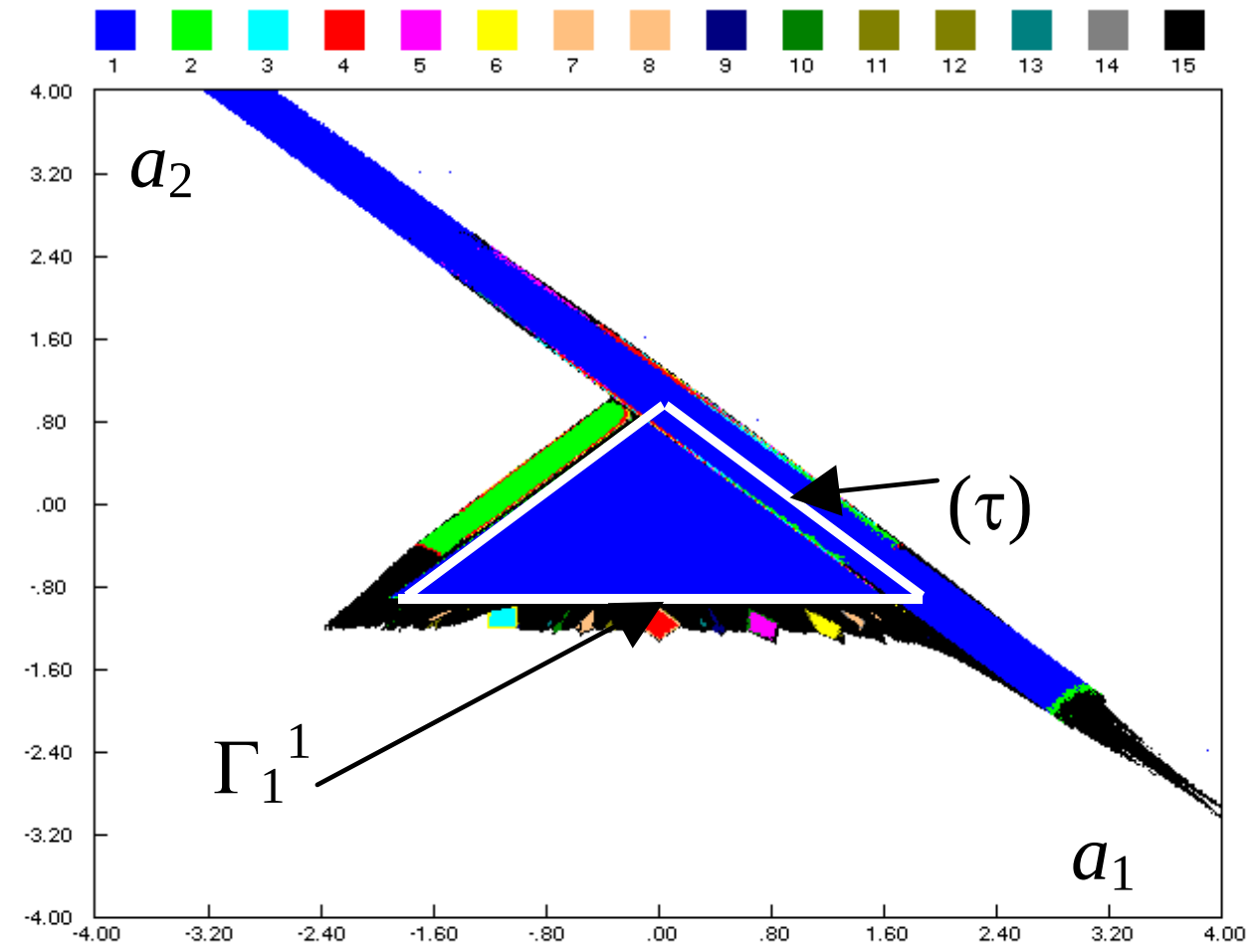

Fig. 9 : Parameter plane (p=1.2 s=3)

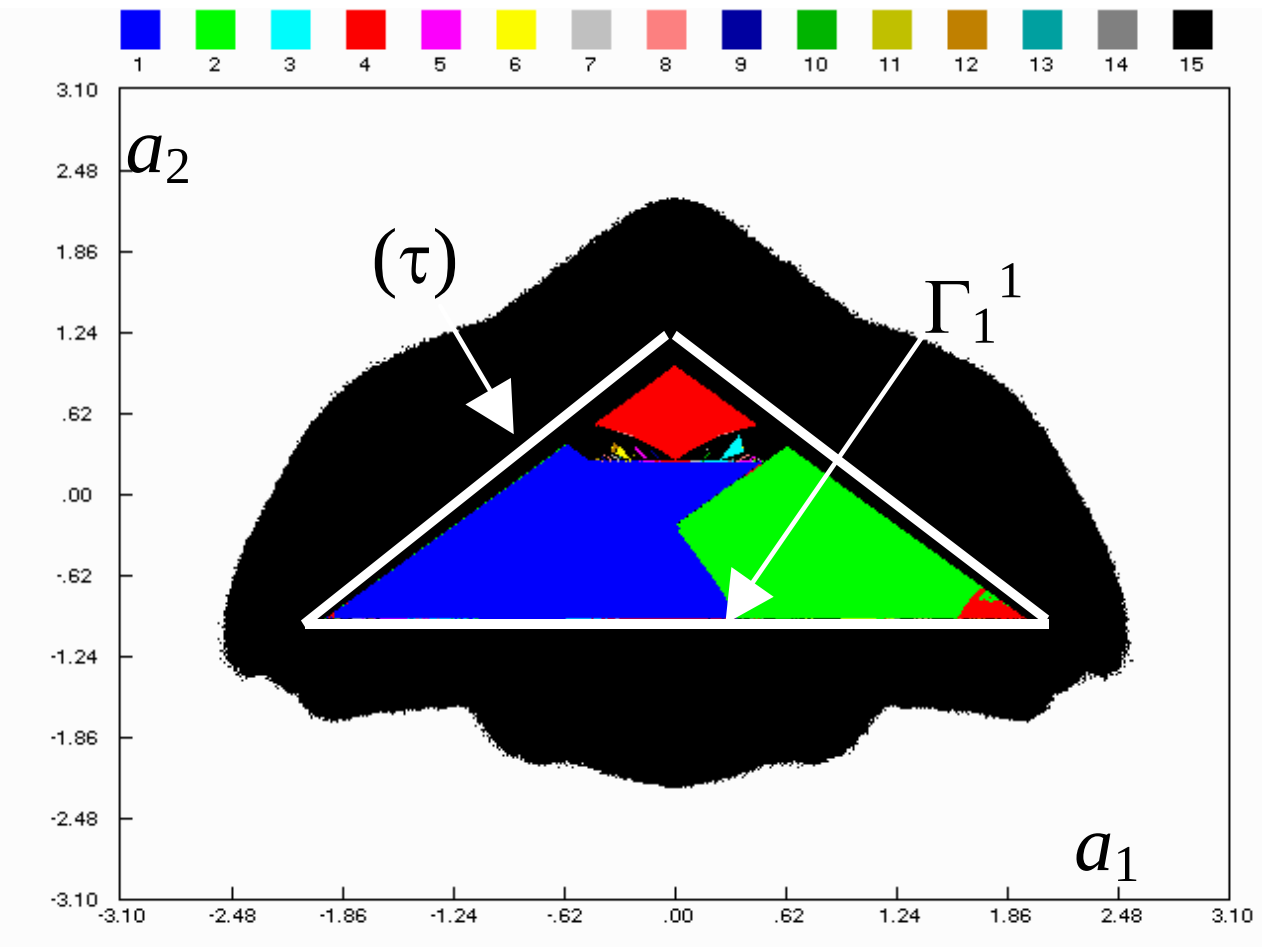

Fig.10 : Parameter plane (p=5 s=0.5)

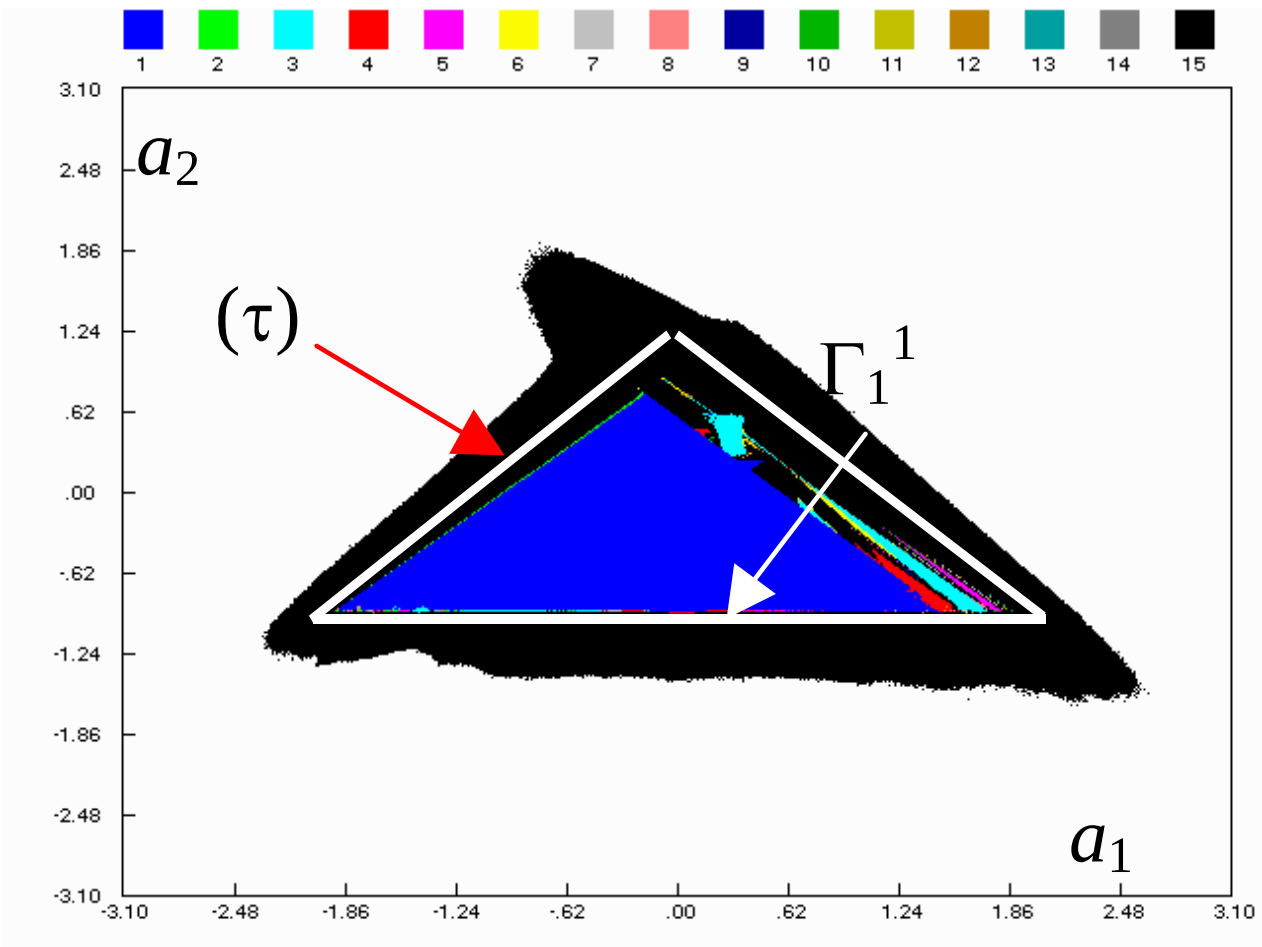

Fig.11 : Parameter plane (p=5 s=1.5)

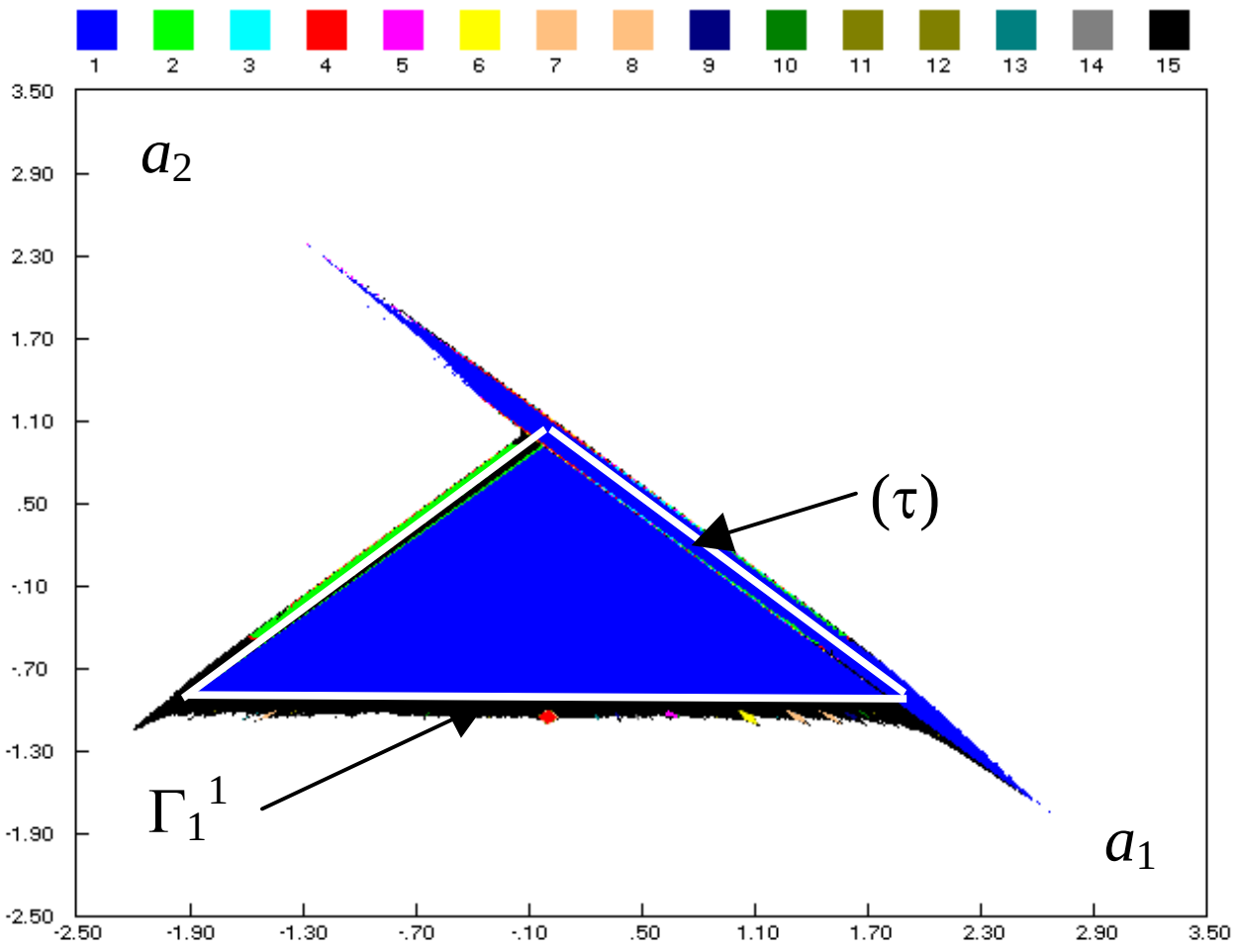

Fig.12 : Parameter plane (p=1.2 s=8)

In the bifurcation diagram of Fig. 14, which is an enlargment of Fig.8, the regions of admissibility of periodic orbits can be seen, with their corresponding rotation numbers (cf. Fig.14, which shows an enlargment of Fig. 8) [Taralova-Roux, 1996]. The parameter plane is divided into regions where all the orbits associated with any rotation number may exist. The structure of the graph is similar to the well-known Arnold tongues structure [Boyland, 1986] [Ogorzalek & Galias, 1991] [Galias Z., 1995] [Kocarev & al., 1994] [Chua & al. , 1990]. The sequence of periodic orbits is separated by one or more chaotic states, as the periodic window is embedded within some chaotic regimes. The system exhibits as a whole a very rich dynamical behaviour.

From the bifurcation diagrams, it can be seen that, by changing the DPCM parameters, we might set the system to qualitatively different types of behaviour, in which case the system settles to a different attractor. The change can be more or less important with regard to the kind of bifurcation which occurs. A complete understanding of such qualitative changes means knowing the topological changes in the structure of the phase portrait at the bifurcation threshold.

## 4.2. ANALYTICAL STUDY OF FIXED POINTS

The fixed points and period 2 orbits of the DPCM system can be analyzed theoretically. The regions where one or more stable fixed points exist are given by the blue color in Fig. 9-12. (dark blue in Fig. 9-11-12 and light blue in Fig.10). From the theoretical study it follows that the map (11) admits at most three fixed points which are called $Q_1$, $Q_2$, and $Q_3$. These fixed points undergo the bifurcations recalled in section 2, but in the way defined in the PWL case. Here, the fixed point $Q_3$ plays a fundamental role. The coordinates of the fixed points satisfy $X = \mathrm{T}X$ and are analytically given by:

$$\begin{cases} \begin{cases} \mathrm{x}_{Q_1} = \mathrm{y}_{Q_1} = -\dfrac{a_1 + a_2}{a_1 + a_2 - 1} \\ \text{when } s - \mathrm{x}_{Q_1} > \dfrac{1}{p} \qquad (17a) \end{cases} \\ \text{or} \\ \begin{cases} \mathrm{x}_{Q_2} = \mathrm{y}_{Q_2} = \dfrac{a_1 + a_2}{a_1 + a_2 - 1} \\ \text{when } s - \mathrm{x}_{Q_2} < -\dfrac{1}{p} \qquad (17b) \end{cases} \\ \text{or} \\ \begin{cases} \mathrm{x}_{Q_3} = \mathrm{y}_{Q_3} = -\dfrac{(a_1 + a_2)ps}{(a_1 + a_2)(1-p) - 1} \\ \text{when } \left| s - \mathrm{x}_{Q_3} \right| < \dfrac{1}{p} \qquad (17c) \end{cases} \end{cases} \qquad (17)$$

The analogous expressions for $Q_1$ and $Q_2$ lead us to consider the symmetry of the regions $\{D_7, D_3\}: |s - x| > \frac{1}{p}$ (cf. Fig. 4) where the fixed points $Q_1$ and $Q_2$ are defined. It is easy to check that the existence of the fixed point $Q_1$ implies the existence of the other fixed point $Q_2$ and vice versa.· In [Macchi & Uhl, 1993] it has been shown that the optimal estimation of the input signal s is achieved for $a_1+a_2=1$. According to (17), the same result can be observed for the fixed point $Q_3$ as its coordinates for this case are $x_{Q3} = y_{Q3} = s$.

Let us consider separately the case when one or several fixed points appear. A particularity of the piece-wise linear case is that one or several fixed points appear/disappear in general in a different way than in the differentiable case. The algorithm for their numerical calculation is given in section 4.4.

From a mathematical point of view the formulas for the analytical calculation (17) can always be applied, but do not necessarily give rise to a fixed point for the map (11). A necessary condition for the resulting fixed point to exist, is that its coordinates must belong to the corresponding region D7, D3 or D5 (cf. Fig.4) which condition is given by the constraints (17a), (17b) and (17c).

Let us consider now the point $Q_3$ . It undergoes a Neïmark-Sacker bifurcation (4) when $a_2$ crosses the value -1. The corresponding curve is denoted by $\Gamma_1^1$ and is shown in Fig. 8-12. On this curve, the determinant of the Jacobian matrix associated with (4) at $Q_3$ is equal to one, with complex conjugated eigenvalues.

The fixed point $Q_3$ is a stable focus above the $\Gamma_1^1$ curve and repulsive focus below. The set of points $(a_1,a_2)$ forming the Neïmark-Sacker bifurcation curve for $Q_3$ is given by line :

$$-2 < a_1 < 2,\ a_2 = -1 \qquad (18)$$

Parametric equations of flip and Neïmark-Sacker (4) bifurcation curves of the $(a_1,a_2)$ plane related to cycles of order $k$ are given by :

$$\begin{cases} X = \mathrm{T}^k(X, a_1, a_2, p, s), & \mathrm{S}_1(X, a_1, a_2, p, s) = -1 \qquad (19) \\ X = \mathrm{T}^k(X, a_1, a_2, p, s), & \mathrm{S}_i(X, a_1, a_2, p, s) = e^{\pm j\varphi},\ i = 1,2 \qquad (20) \end{cases}$$

$X$ being the coordinates vector of a cycle point and $S_i$ being a multiplier of the k-cycle. Bifurcation curves are numerically obtained in the $(a_1,a_2)$ plane using Newton method.

## 4.3. GRAPHICAL STUDY OF PERIOD TWO POINTS

Writing the map $T^2$ in the following form,

$$T^2(X) - X = \begin{cases} \delta(x_k, y_k) \\ \bar{\delta}(x_k, y_k) \end{cases} \tag{21}$$

the periodic orbits of period two satisfy the conditions

$$\delta(x_k, y_k) = \bar{\delta}(x_k, y_k) = 0 \tag{22}$$

Let us represent these conditions graphically in the phase plane for the particular parameter values for which (22) is verified (Fig. 13).

The period-2 orbit exists when the two curves (22) intersect each other. For the example considered, they coincide over a line segment. It can be proved that the slope of this segment is equal to -1. Therefore an infinity of period two points exists for all starting conditions on the red-blue overlapping segment (recall that $x_{n+1} = y_n$). In this particular case, bifurcation occurs with multipliers equal to one, and, because of the infinity of solutions, is called a “degenerate” solution. Each starting condition taken in the overlapping segment corresponds to a different period two orbit. After the bifurcation, the period-2 orbits disappear, since the condition (22) is not satisfied anymore. The same phenomenon (infinity of appearing periodic orbits) can be generalized for higher order periodic orbits.

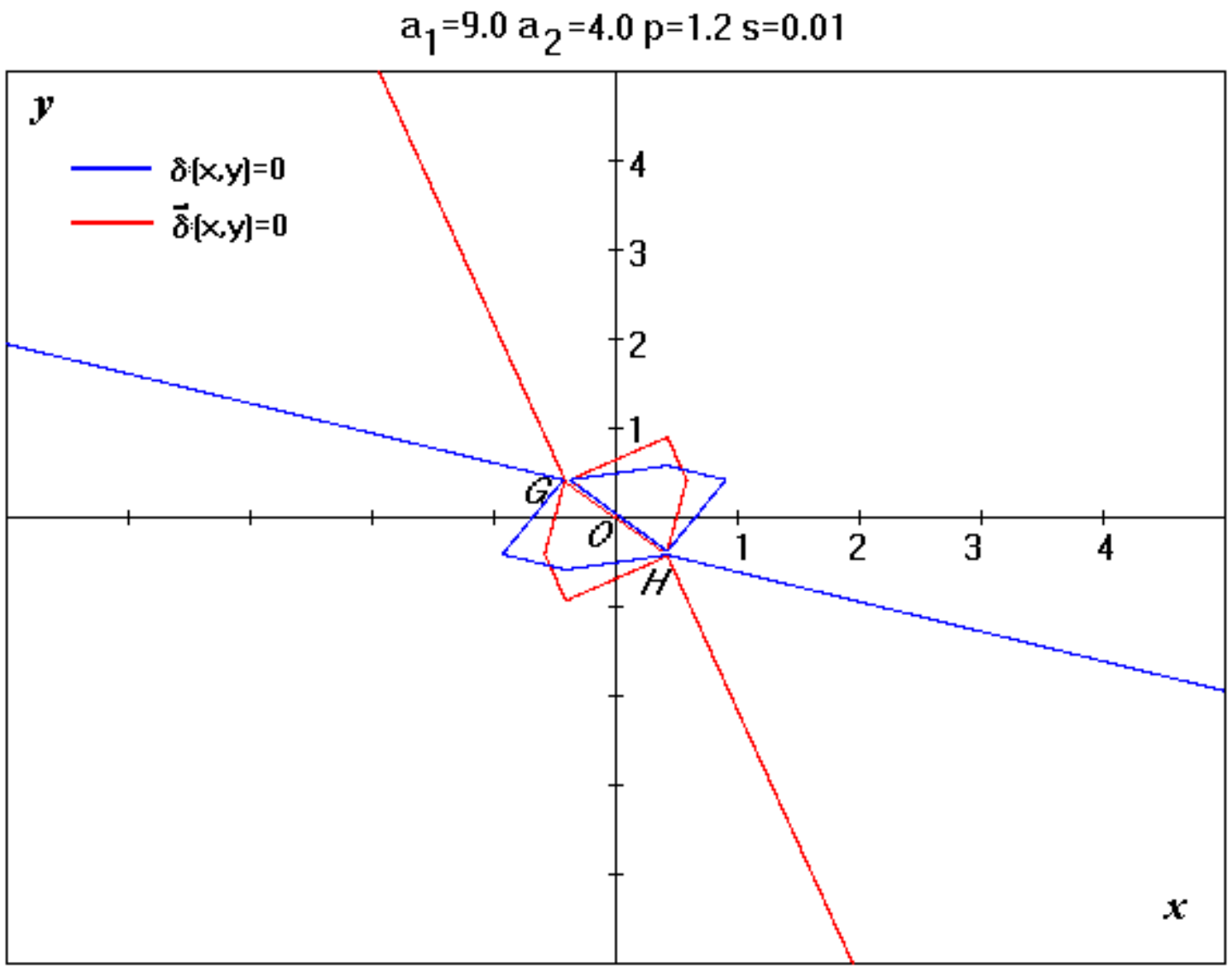

Fig.13 : Degenerated solutions with multiplier equal to one

## 4.4. NUMERICAL STUDY AND ALGORITHM FOR THE BIFURCATION CURVES CALCULATION

Numerical calculation techniques are used to examine the behaviour of fixed and periodic points as a parameter is varying and to follow their bifurcations in the ($a_1$, $a_2$) parameter plane. The bifurcation structure of the map (11) is established with the help of numerical methods. Indeed, the analytical study can be done easily only for the fixed points and the period 2 orbits of the map (11). For a cycle of order k, when k≥2, many analytical difficulties arise related to the exponentially growing number of the different determinations regions for the map $T^k$, k = 1, 2, …, which require the use of numerical methods.

Looking back at the bifurcation diagrams a question arises : why is it necessary to calculate these bifurcation curves if we have already undertaken the parameter plane scanning? The answer is that bifurcation curves give more precise and detailed information about the areas with different kinds of behaviour, this is particularly important when these curves overlap. In terms of phase space interpretation, it means that two or more different stable types of behaviour may coexist, and the resulting phenomenon is known as *multistability*. This is one of the aspects of the richness of the non-linear systems.

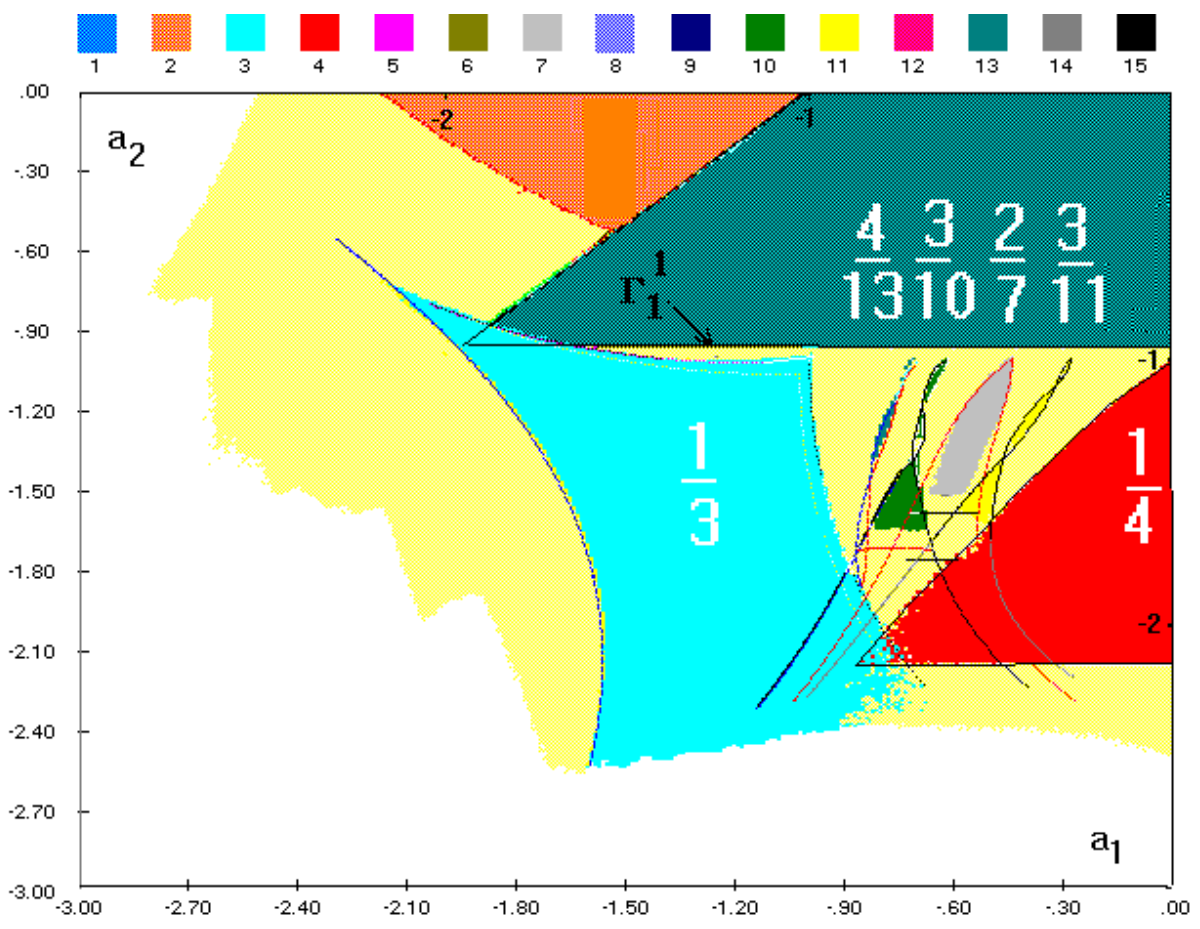

Fig. 14 : Enlargment of Fig. 8 with the corresponding rotation numbers (p=1.2 s=0.5).

In the differentiable case, in order to obtain the bifurcation curves in the parameter plane, we have to calculate the bifurcation curve $G(a_1, a_2) = 0$ satisfying

$$\begin{cases} x = f^k(x,y,a_1,a_2) \\ y = g^k(x,y,a_1,a_2) \\ S(x,y,a_1,a_2) = 1 \end{cases} \qquad (23)$$

This is quite natural, since a state bifurcates (e.g. from stable to unstable) when its multipliers cross the unit circle. On the other hand, for a piece-wise linear map, cycles can appear or disappear without having their multipliers crossing through +1, as has been mentioned at the beginning of section 3, i.e. a criterion of the type (23) would not be useful to calculate the bifurcation curves.

In our PWL case the aim is the same as before , i.e. to find

$$G(a_1, a_2) = 0 \qquad (24)$$

but satisfying

$$\begin{cases} x = f^k(x,y,a_1,a_2) \\ y = g^k(x,y,a_1,a_2) \\ (x,y) \in \mathrm{E_D} \end{cases} \qquad (25)$$

where $\mathrm{E_D}$ denotes the set defined by (11a) and (11b). This condition replaces the standard one because in the PWL map periodic orbits arise when a point of the cycle hits the lines $\mathrm{E_D}$. A numerical algorithm has been developed based on the conditions (25), and the obtained bifurcation curves with the corresponding rotation numbers are presented in Fig. 14. Note that the color scheme has been changed with respect to Fig. 8 in order to achieve better contrast; the bifurcation curves (in black) correspond to orbits with different period and rotation number (4/13; 3/10; 2/7; 3/11), embedded in the chaotic regime (in light yellow).

Such “saussage-like shape” of the bifurcation curves has been already observed for digital filters [Ogorzalek, 1992].

The ordering of the rotation numbers is similar to that of the bifurcation structure “boxes in files” (also called Arnold tongues structure) for the one dimensional case arising at the Neïmark-Saker bifurcation, the frontier of each “box” being a bifurcation curve (see [Mira, 1987] for more details). This ordering strictly depends on the degree of complexity $D_c$ of the box and is a function of its rotation number. Two neighbouring boxes with rotation numbers *b/c* and *d/e* which have the same degree of complexity $D_c$ will satisfy the equation $cd - be = \pm 1$. All boxes with rotation numbers *1/k* have degree of complexity $D_c$ =1. Between any two boxes *1/k* and *1/(k+1)* there exists an infinite file of boxes with degree of complexity $D_c$ =2. In the example of Fig. 14, the sequence of rotation numbers 4/13; 3/10; 2/7; 3/11 is part of a file with degree of complexity $D_c$ =2, and is embedded between the boxes 1/3 and 1/4 with degree of complexity $D_c$ =1. In general, between any two boxes with degree of complexity $D_c$ there is an infinite file of boxes with degree of complexity $D_c$ +1 and so on. This bifurcation structure is fractal; therefore stable orbits with any period and any rational rotation number could be obtained and their location in the parameter plane could be predicted.

## 5. STUDIES IN THE PHASE PLANE. BASINS.

Once the attractor(s) and the type of each are known, one would like to associate each attractor with the set of all starting conditions leading to it; i.e. the basin of attraction. This notion concerns the analysis of the system behaviour in the phase plane via the *phase portrait*. For a given point of the parameter plane ($a_1$, $a_2$), several different stable behaviours could coexist, this is one example of the richness of non-linear systems. The state vector trajectory depends on its initial conditions; furthermore, realizations starting from nearly the same initial condition may result in (extremely) different final states. The explanation of this phenomenon is related to the basin boundaries and will be given below. This notion is related to the robustness of the phase portrait against sufficiently small perturbations of the parameters.

In this section, the first and second paragraphs are devoted to the description of attractors issued from Neïmark-Sacker bifurcations and the third one to comments about basins and their boundaries.

### 5.1. NEÏMARK-SACKER BIFURCATION OF A FIXED POINT

The study of the different attractors reveals extremely interesting phenomena associated with the *piece-wise* linear structure of the quantizer characteristic. One of them concerns the bifurcation (4). In PWL maps, the case related to an irrational rotation number $\mu$ does not give rise to a differentiable ICC but to an ICC, which can be called a *Weakly Chaotic Ring* (WCR) as in [Mira & al. , 1996], the Lyapunov exponent of which is slightly positive. When parameters are varied, from the bifurcation (4), the Lyapunov exponent increases and the WCR leads to a stronger chaotic attractor. Let us explain the specific phenomenon in the PWL case.

The most important difference with respect to the differentiable case is the appearance of an ICC in the vicinity of the Neïmark-Sacker bifurcation (Fig. 15a) with a large amplitude i.e. far away from the destabilized focus. This sudden change does not occur in the differentiable case, when the invariant closed curve grows continuously around the focus point. The peculiarity is based on the same phenomenon that has already been pointed out above: the fact that at least two points of the attractor belong to two different regions. Physically this phenomenon is expressed by large amplitude oscillations in the immediate neighborhood of the focus-type steady state. This drastic change in the system behaviour affects the good functioning of the system, and is specific to the piece-wise linear character of the quantizer.

If we consider a region Di in the phase plane, where each point has only one rank-one preimage, thus locally we are in presence of a typical linear case. In the linear case, the only possible singularities are the fixed points (attractive or repulsive) and the closed invariant curves resulting from complex multipliers with modulus ρ=1 (center fixed point). In the PWL case, the ICC arises at a contact with the lines defined by (11a)(11b) (LCa and LCb limiting Z3 are images of the lines (11a)). This implies the ICC to be nondifferentiable at at least one point. But the image of an angular point is another angular point and the ICC is invariant under ***T***. As the accumulation of points is everywhere dense, the ICC is constituted by an infinity of angular points. It seems to be a fractal curve in the sense that the curve is not differentiable in any point. Moreover, this property of nondifferentiability is due to the crossing of the ICC through $LC_{-1}$ curve. This is the reason why we call it a WCR, as in [Mira & al. , 1996]. In Fig.15a, the WCR has just appeared (Neïmark-Sacker bifurcation arises for $a_2$=-1.0), touching the lines (11a-b).

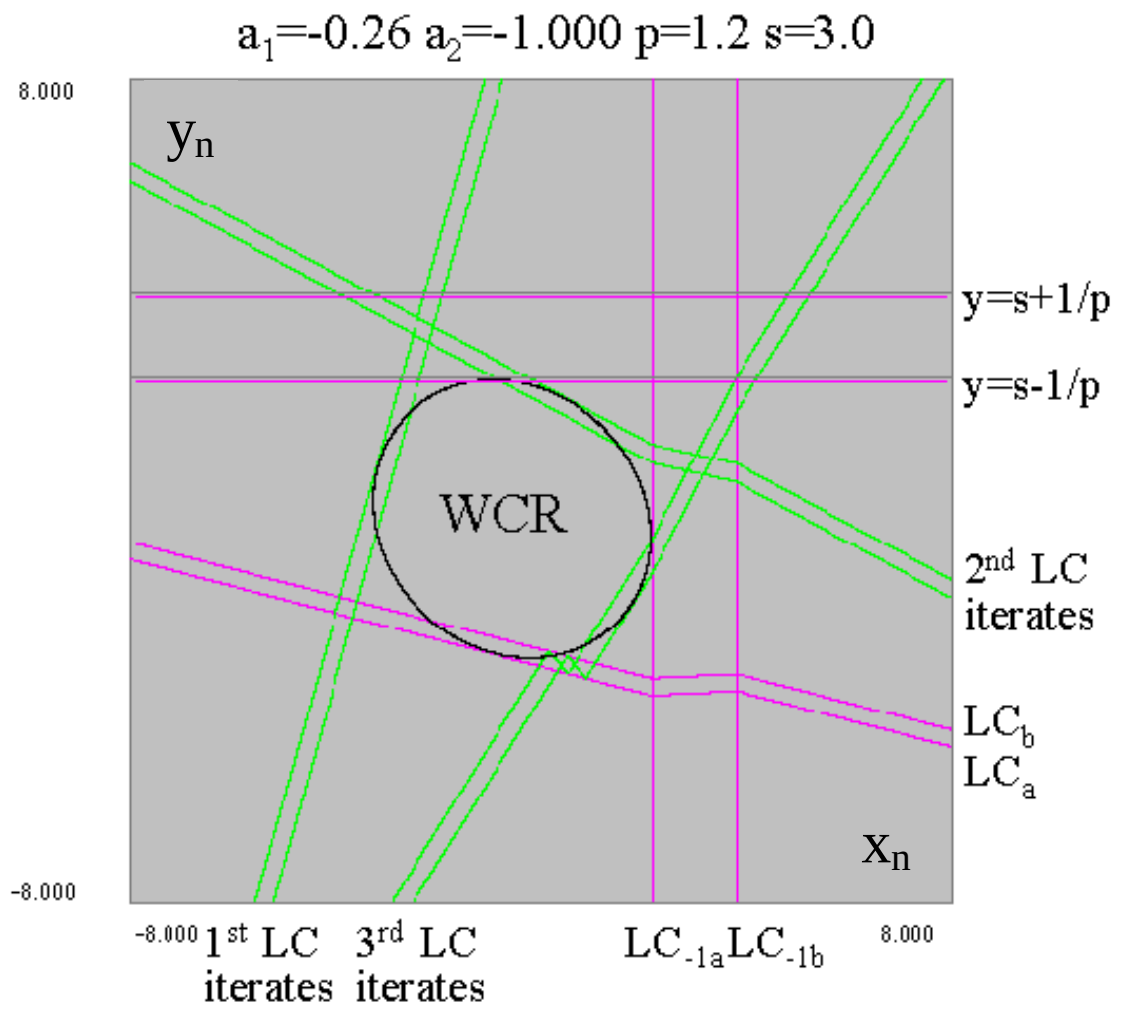

Fig. 15a : Invariant closed curve in the vicinity of the Neïmark-Sacker bifurcation, $a_2$ = -1.00000001

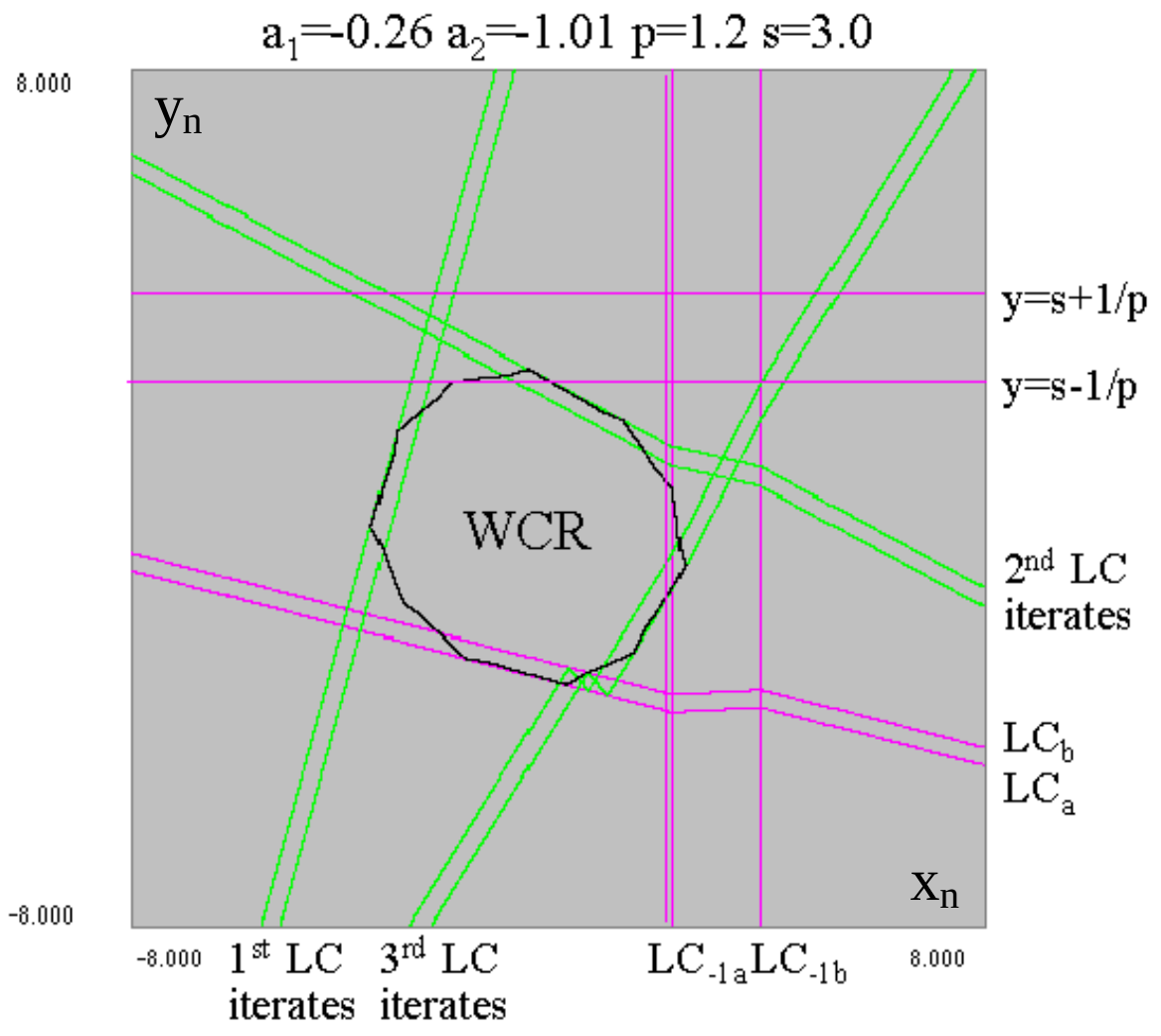

Fig. 15b : When tuning one parameter ($a_2$), the nondifferentiability of the WCR clearly appears.

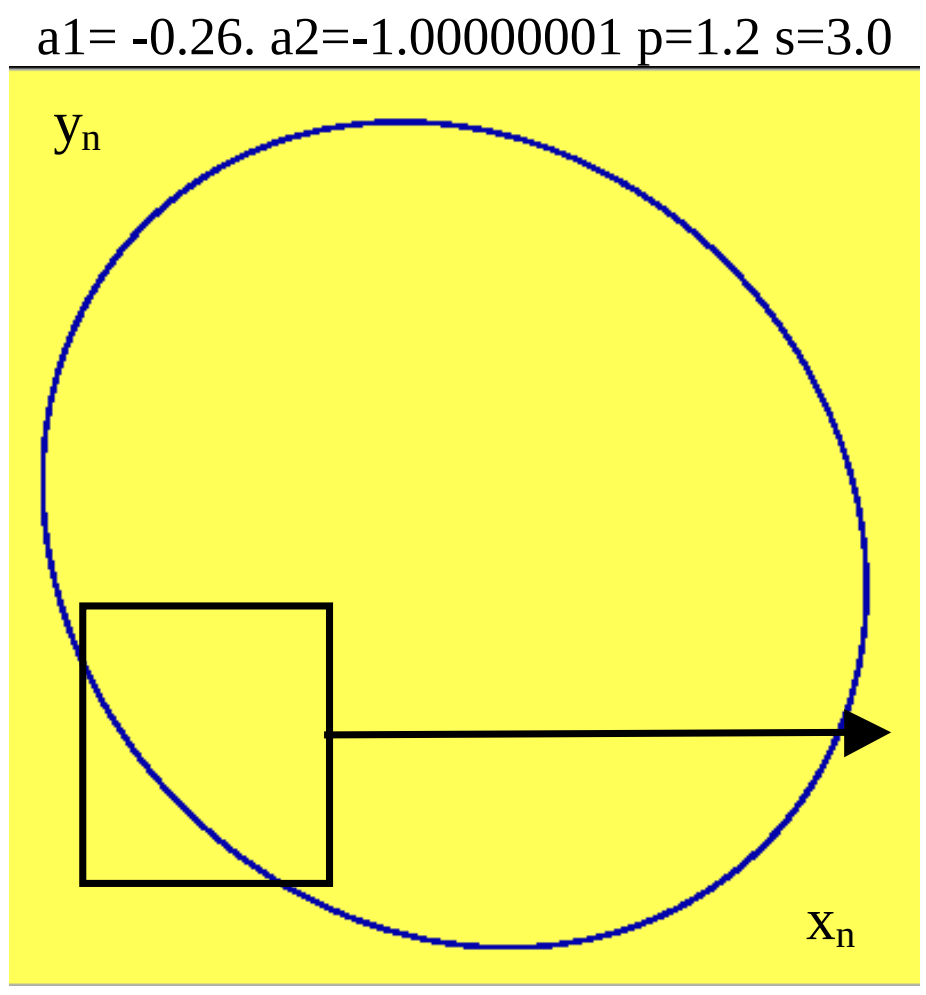

Fig. 15c : The Weakly Chaotic Ring (WCR)

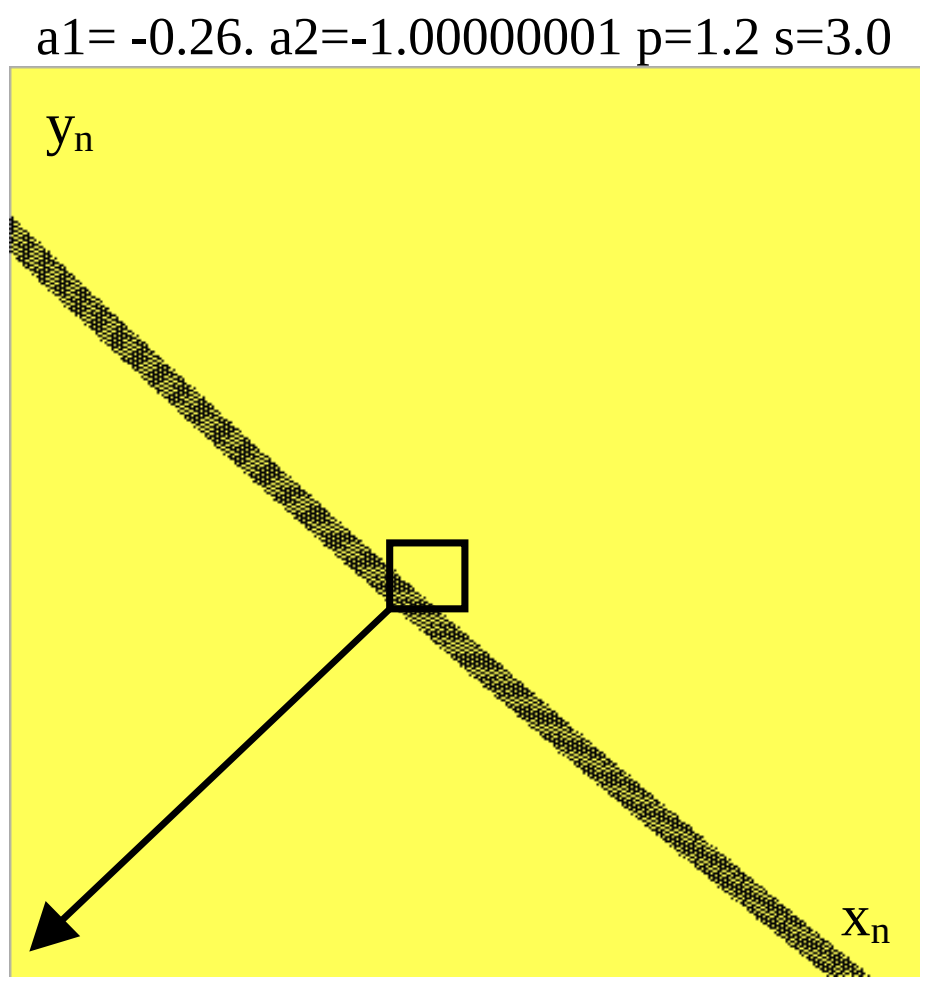

Fig. 15d : Enlargment of Fig.15c

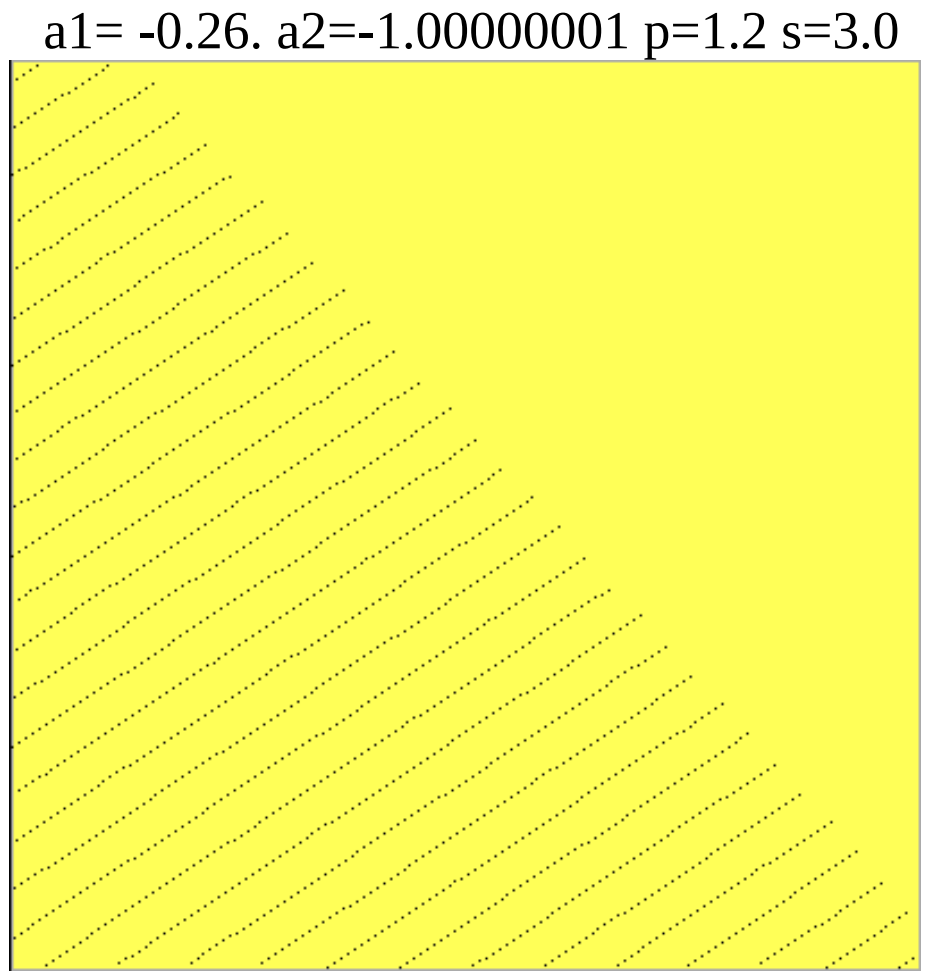

Fig. 15e : Enlargment of Fig.15d

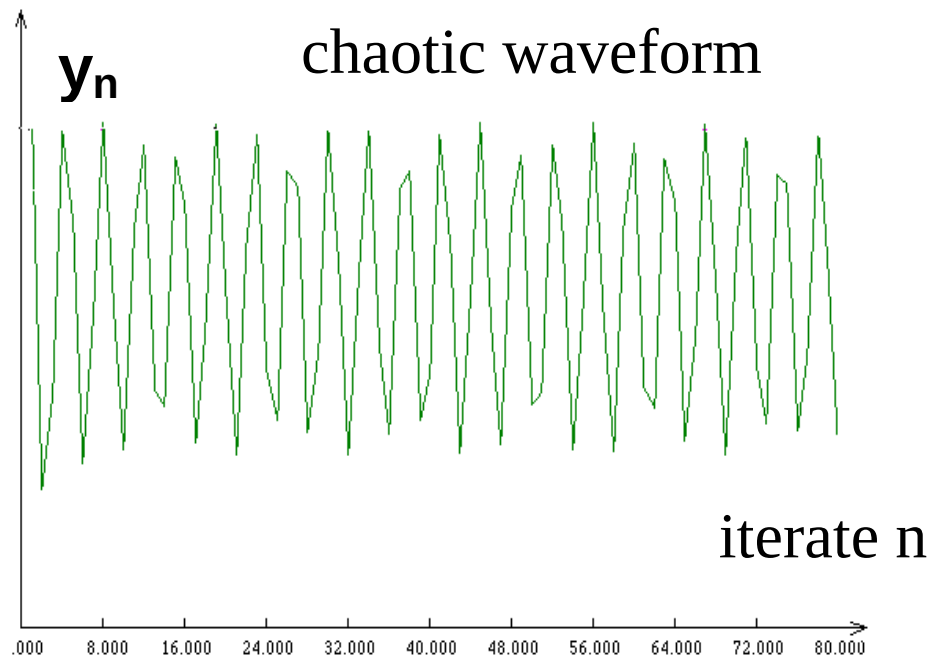

Fig. 15f : Time representation of the WCR
($a_1$ = -.26, $a_2$ = - $a_2$ = 1.00000001, $p$ = 1.2, $s$ = 3)

A WCR can only exist in a region where at least two different determinations are needed. In the considered case, a WCR intersects a $Z_3$ zone. Being an attractor, it is delimited by the critical lines and their iterates [Mira & al., 1996]. Since the analytical expressions of the critical lines are known, we are able to predict the maximum amplitude of the chaotic oscillations when the WCR becomes a stronger chaotic attractor.

In the Figs.15c-15e, three consecutive enlargments of the WCR are presented, which reveal its typical characteristic of self-similarity at different scales. The fractal property and the self-similarity of the attractor in Fig. 15a may let us think it is a chaotic attractor, and not a simple closed invariant curve. One of the characteristics of the chaotic attractor is that for two nearly identical conditions the two resulting motions may diverge at an exponential rate. Of course if the starting conditions were precisely the same, then the deterministic nature of the equation guarantees that the motions are identical for all time. But since some uncertainty in the starting condition is inevitable with real physical systems, the divergence of nominally identical motions cannot be avoided in the chaotic regime. Fig. 15f shows the corresponding chaotic waveform in the (n, $y_n$) plane.

## 5.2. NEÏMARK-SACKER BIFURCATION OF CYCLES

Let us now consider the Neïmark-Sacker bifurcation for period-k cycles. As mentioned previously, when $a_2$=-1 a Neïmark-Sacker bifurcation (4) occurs. Unlike the differentiable case, here an infinite number of elliptic orbits can appear [Taralova-Roux & Fournier-Prunaret, 1996b], with periodicity corresponding to the periods of the different stable orbits before the bifurcation. This situation is analogous to the one which occurs in the model of analogue-to-digital and digital-to-analogue converters [Feely & Chua, 1991] [Feely & Fitzgerald, 1996] [Feely & al., 2000].

The case $s$=0 is particularly studied, since according to the use of the DPCM when the input is zero, the decoded output should be zero as well; in fact, our results demonstrate that the inherent nonlinearity of the quantizer may provoke unwanted oscillations at the output whose shape in the time domain is similar to that of Fig.15f even with the simple piece-wise linear model (Fig. 2b). The parameter p is chosen large enough (p=60), because in this case the model is very close to the model given by Fig.2a, (staircase characteristic); the behaviour of the system is similar to that of a staircase characteristic with two steps (quantizer with one bit).

For the chosen parameter values, in the phase plane ($x_n$,$y_n$), we observe elliptic-like orbits, since $x_n$ and $y_n$ correspond to two consecutive estimations of the input signal ($x_n$ is the memorised $y_n$ value after one iteration).The infinite number of such elliptic orbits at the bifurcation is due to the fact that each different initial condition inside the region in which the map is linear gives rise to a different elliptic trajectory, with the same periodicity (locally we have center-type orbits). The largest ellipse is the one in contact with the lines (11a-b). An infinity of such elliptic orbits exists for the same periodic point of periodicity $k$. Since the phenomenon of elliptic-like behaviour takes place for all periodic points of order $k$ ($k$=1,2,8,24 in Fig. 24), the starting conditions where the system is initialized determine then which discrete elliptic region will capture the system trajectories. The set of all elliptic regions is enclosed inside a trapping region, also called an absorbing area. This area captures all system trajectories after a finite number of iterations. In our case, the borders of the absorbing area are completely defined by the critical curve segments and their first iterates.

Just after the bifurcation, these elliptic orbits give rise to sets of cyclic WCR with periodicity corresponding to that before the bifurcation (in the considered case (Fig. 25) we observe two chaotic rings of period 1, one set with period 2, one of period 8 and one of period 24).

Again, the borders of the domains of existence of the set of chaotic rings are completely defined by the critical curve segments and their first iterates (cf. Fig. 25). The WCR coexist until $a_2$=-1.003; after this value, the chaotic sets merge into a unique chaotic attractor which disappears in turn by a classical bifurcation (contact of the chaotic attractor with its own basin). This latter phenomenon has also been observed in the model of a sigma-delta modulator [Feely & al., 2000].

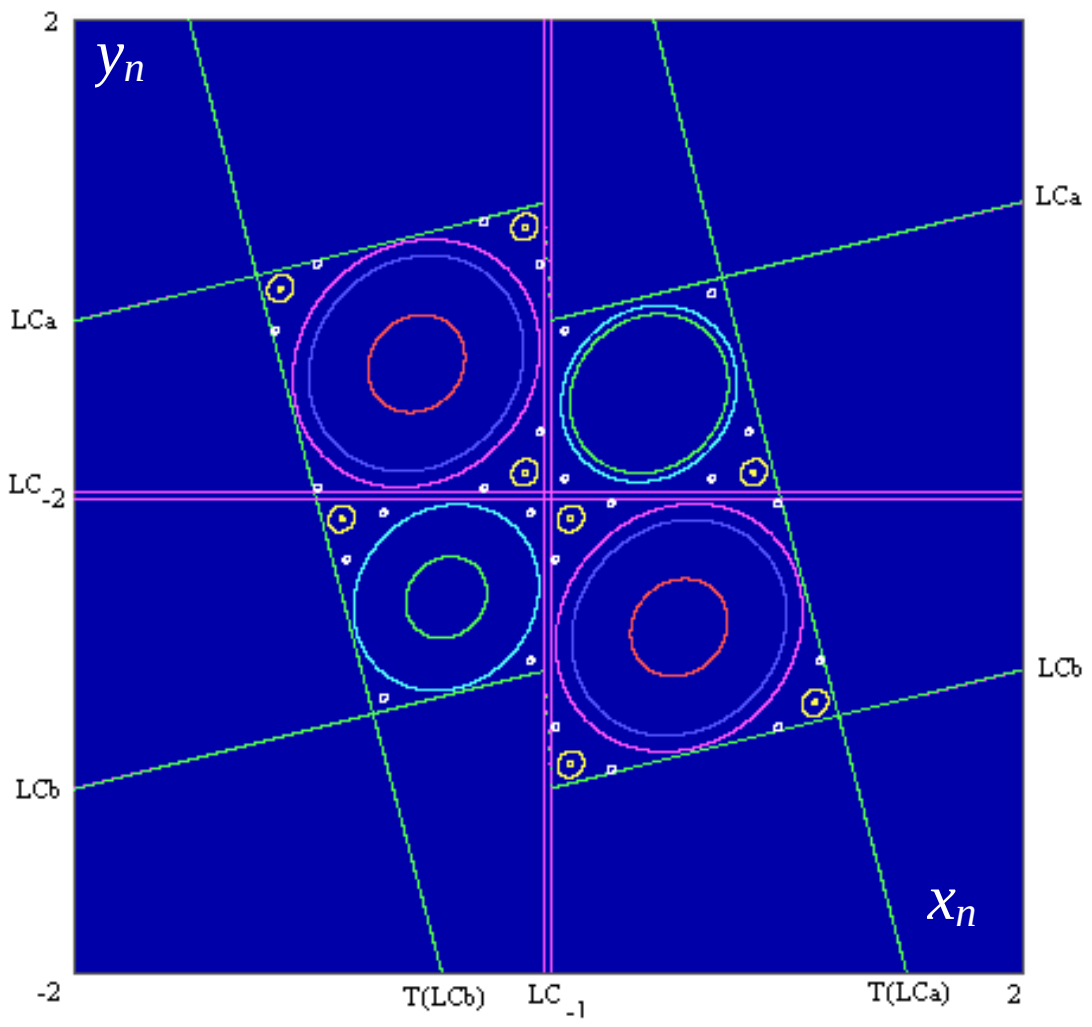

Fig. 24 : $a_1$=0.25 $a_2$=-1.00 p=60 s=0.0. Two periodic orbits of order 1, one of period 2, one of period 8 and one of period 24 can be observed.

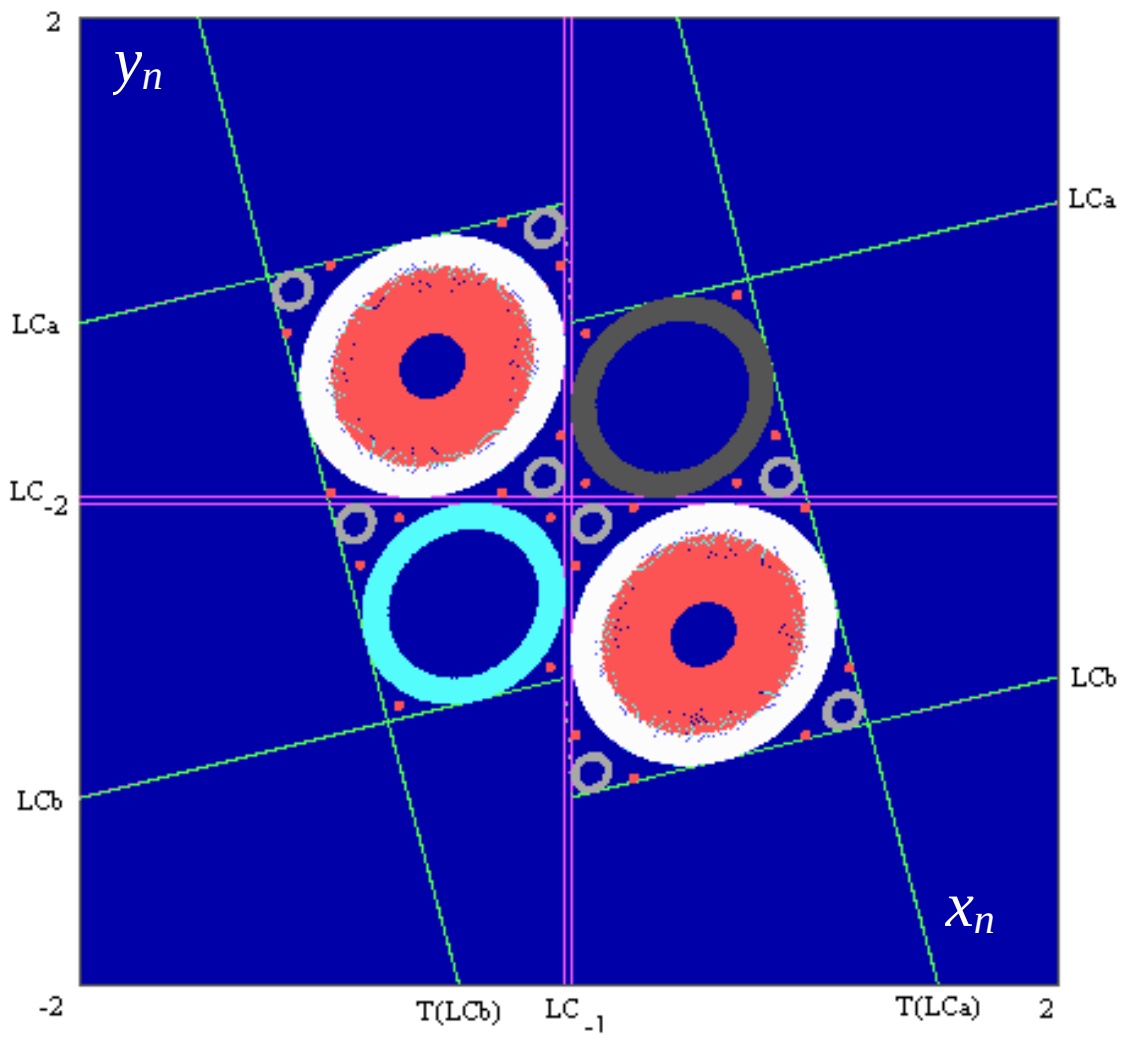

Fig. 25 : $a_1$=0.25 $a_2$=-1.001 p=60 s=0.0, after the Neïmark-Sacker bifurcation.

## 5.3. MULTISTABILITY AND BASINS EVOLUTION

In general multistability is intrinsic to many nonlinear systems of interest to electronic and system engineers. For an appropriate choice of the operating conditions in a DPCM transmission system coexistence of up to four attractors has been observed. The existence of different attracting steady states destroys the global stability of the operating steady state, particularly if their basins are strongly intermingled. We then study in the phase plane the evolution of the attractors and that of their basins and demonstrate the applicability of the critical lines tool in the case of DPCM map.

Detailed phase plane study has been performed below the line of Neïmark bifurcation (boxes-in-files bifurcation structure as the one observed on Fig. 14) for constant $a_2$ and variable $a_1$, and vice versa in order to analyse separately the influence of the predictor parameters on the system behaviour. Unlike the previous studies of similar systems [Ogorzalek, 1991], chaotic behaviour is very often observed due to the boxes' overlap. In our next examples, chaotic attractors have been obtained by numerical simulations. The figures show that up to four distinct attractors may coexist, a result which has not been observed in previous differentiable models of the system (only three coexisting attractors have been observed). It is important to analyse whether this functioning mode is inauspicious or not for the system, as each of these modes is strongly stable and does not provoke in itself a bad functioning.

The width of the corresponding box is important as well, because the existence of a given periodic point-type attractor is guaranteed inside the tongue, but small changes in the predictor parameters may change the functioning point moving it into a region with an unpredictable or chaotic behaviour. In this sense regions with several coexisting attractors in general have to be avoided, because higher order periodic points imply finer Arnold tongues bifurcation curves, and are therefore prone to switch to an unwanted dynamical behaviour with higher probability under slight parameter changes.

In order to better understand this phenomenon, let us consider a small window inside the Arnold tongues (recall that the Neïmark-Sacker bifurcation occurred at $a_2$=-1). The case of two distinct attractors, a stable period-11 focus and a stable period-26 node, is shown in Fig.16a; $a_2$ is fixed slightly below $\Gamma_1^1$, so slightly below the stability triangle (section 4.1) and $a_1$ is varied. Each initial condition plotted in red corresponds to trajectories converging towards the period-11 focus, and each initial condition taken from the region in grey gives rise to trajectories converging towards the period-26 node. So the red domain is the basin of the period-11 focus and the grey one is that of the period-26 node. Between the basins of attraction there is a separator curve, and it is clear that two rather close starts straddle this separator and that their evolution is uncertain. This problem can spread to the whole phase plane after a small parameter tuning (Fig. 16b) that makes the basin's boundary become uncertain in a larger part of the phase plane : the basin of the period-11 focus is spread in a countable number of nonconnected components all over the basin of the period-26 node. It is difficult to predict even quantitatively whether the trajectories of the system will be captured by the period-26 node or by the period-11 focus, because all initial conditions are tangible numerically with finite precision. In this case the boundary is said to be *fuzzy*. This terminology is introduced in [Mira & al., 1987]; fuzzy is equivalent to fractal. In the considered case both cycles are stable, but the problem can become more intricate when the basin of the attractor is disconnected with divergent regions inside [Gicquel, 1995].

When the parameter $a_1$ increases, a forth multistability case (Fig.17) appears. As already mentioned, this fact implies that in the parameter plane four boxes overlap.

Besides the stable periodic points, many other unstable periodic points exist as well. Saddle points have the peculiarity that in many cases their stable manifolds (associated with the multiplier $|S|<1$) limit the basins of attraction [Mira & al., 1996] (Fig.18). On the contrary, the unstable manifolds (associated with the multiplier $|S|>1$) converge towards the attractors. In the case under consideration, we can observe a chaotic transient due to the existence of a strange repellor (Fig. 19) which, under a slight parameter change, (Fig.20a & 20b) becomes a strange (chaotic) attractor when no stable cycle exists. From Fig. 20a another interesting feature of the critical lines tool is emphasized: its iterates (in green) limit the strange attractor (in red). When the parameters are varied, this attractor undergoes quantitative changes in its form and shape and coexists with other stable periodic orbits (Fig.21).

Similar complex phenomena occur on the symmetric region for positive $a_1$ values (Fig.23); as an example, a fuzzy frontier in the case of coexisting period-17 focus and 21 node is given. Although the cycle orders are different from those of Fig.17, if we compare the two figures we can see that the basin structure remains qualitatively the same.

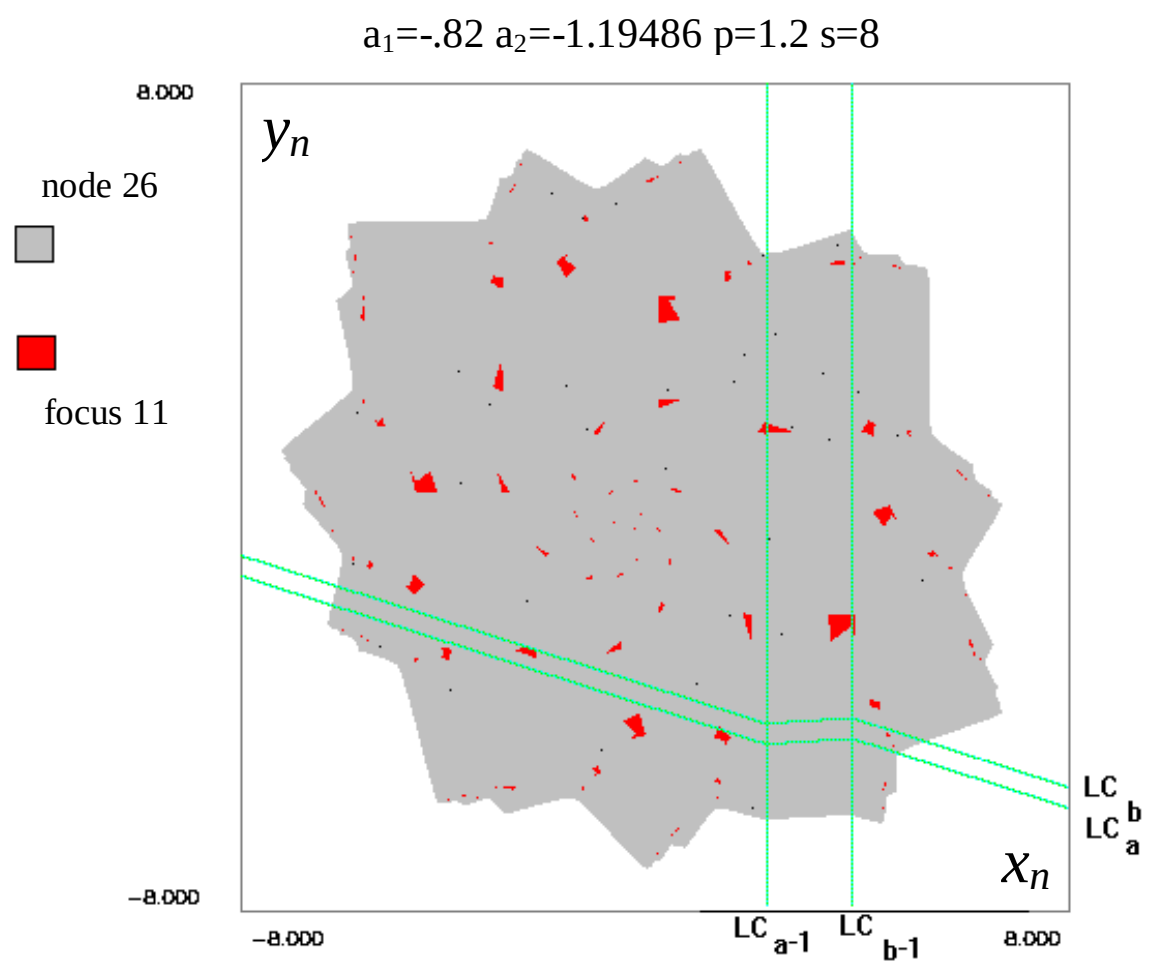

Fig. 16a : Period-11 focus and period-26 node coexist.

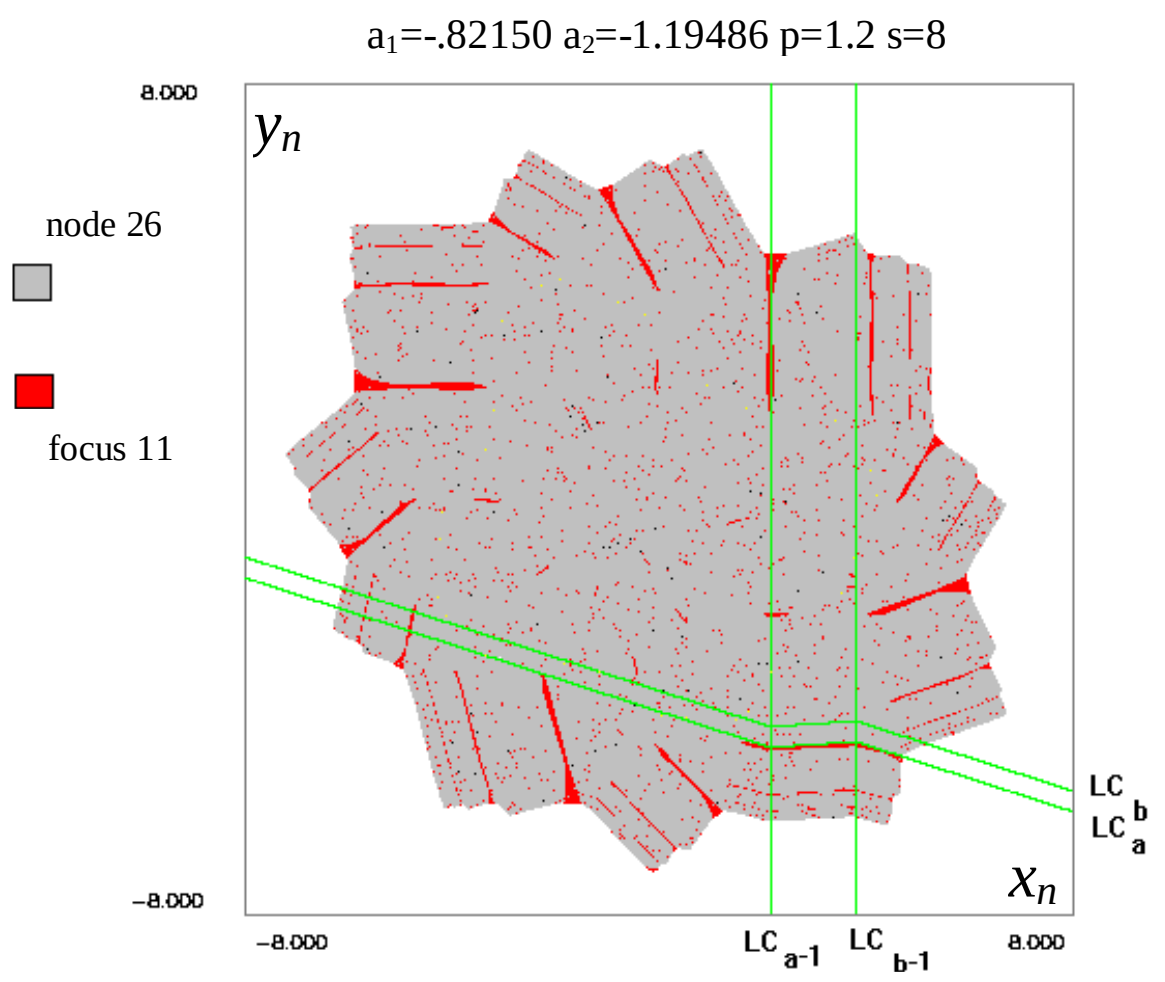

Fig.16b : After slightly tuning the parameter $a_1$, the boundary between the basins becomes fuzzy.

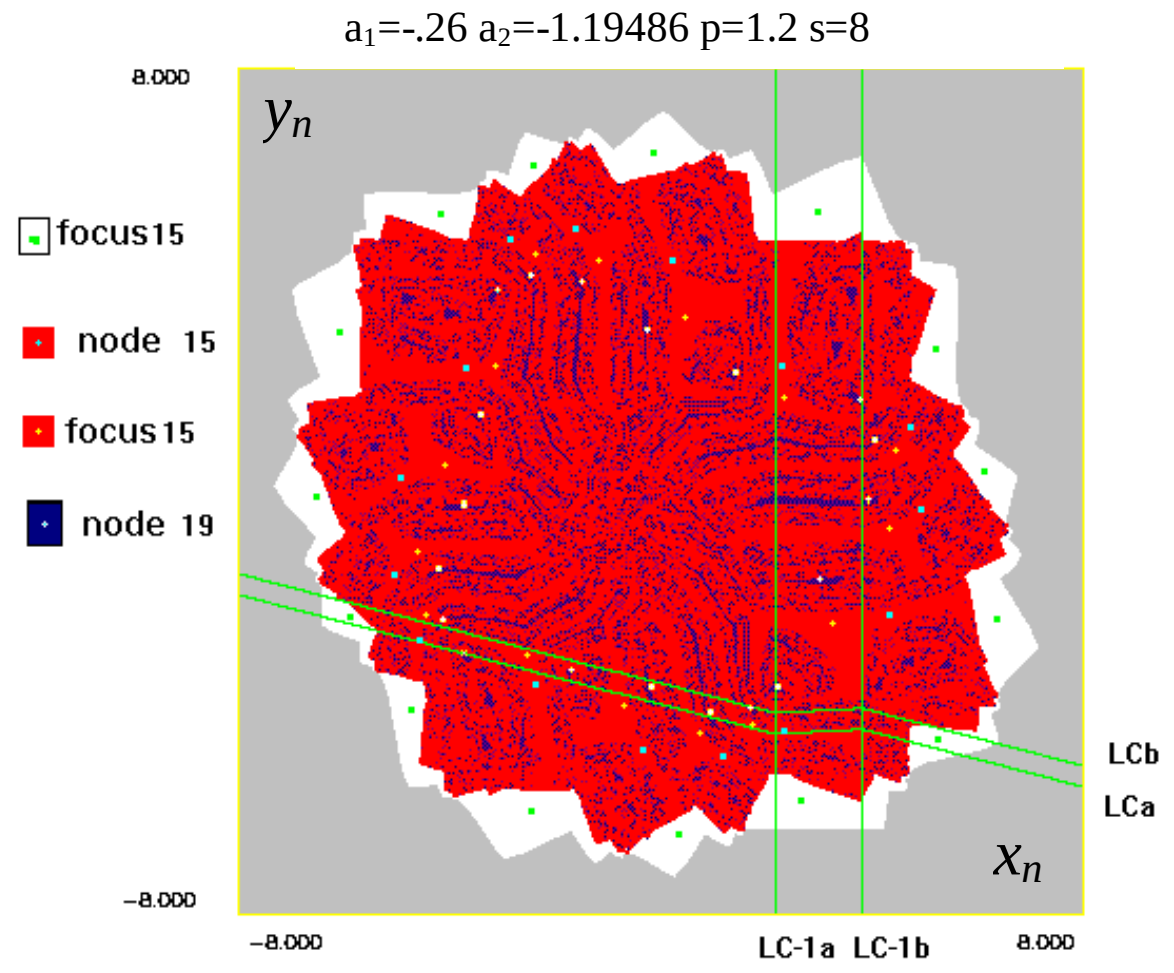

Fig. 17 : Four attractors coexist, some of the basins are separated by a fuzzy boundary.

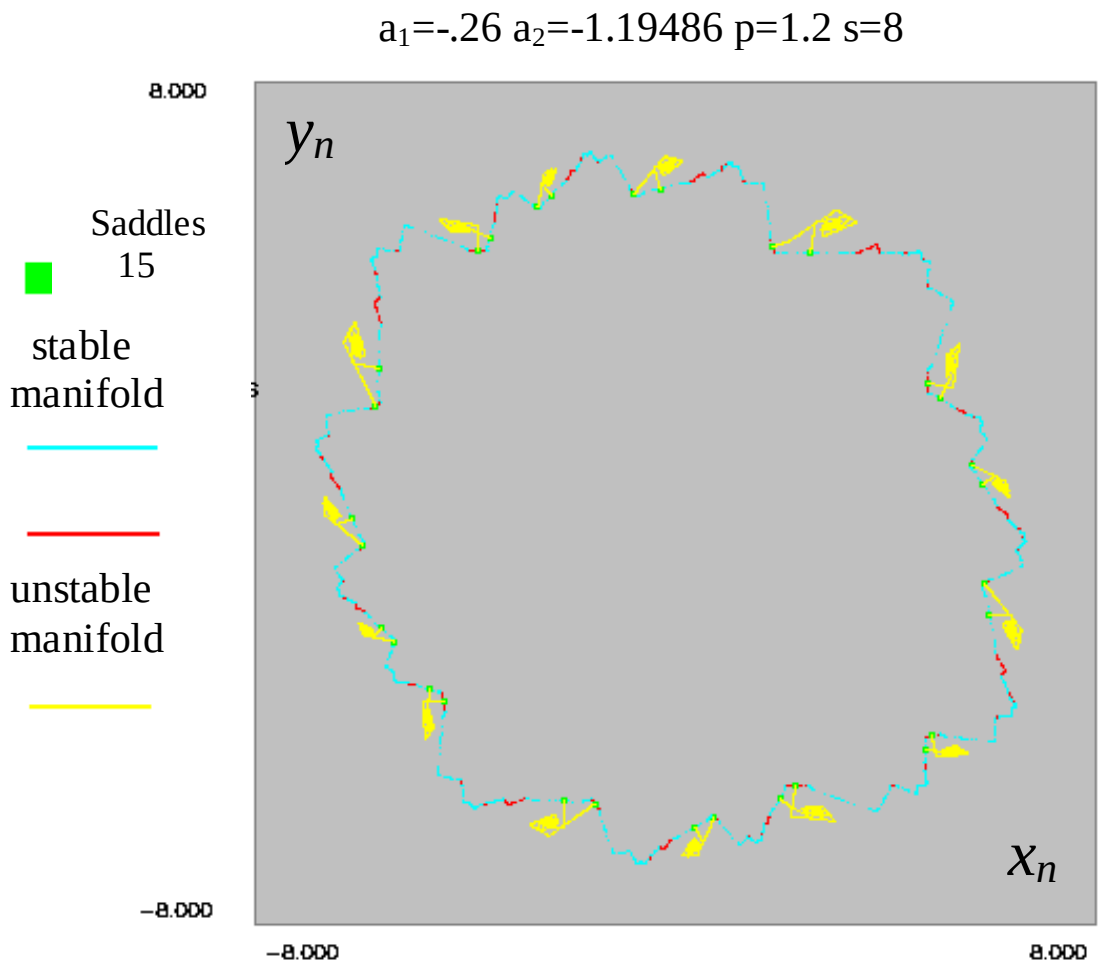

Fig. 18 : The basin frontier is formed by the stable manifolds (cyan and red colours) of the two period-15 saddles. The unstable manifold is plotted in yellow.

$a_1$=-26 $a_2$=-1.19486 p=1.2 s=8

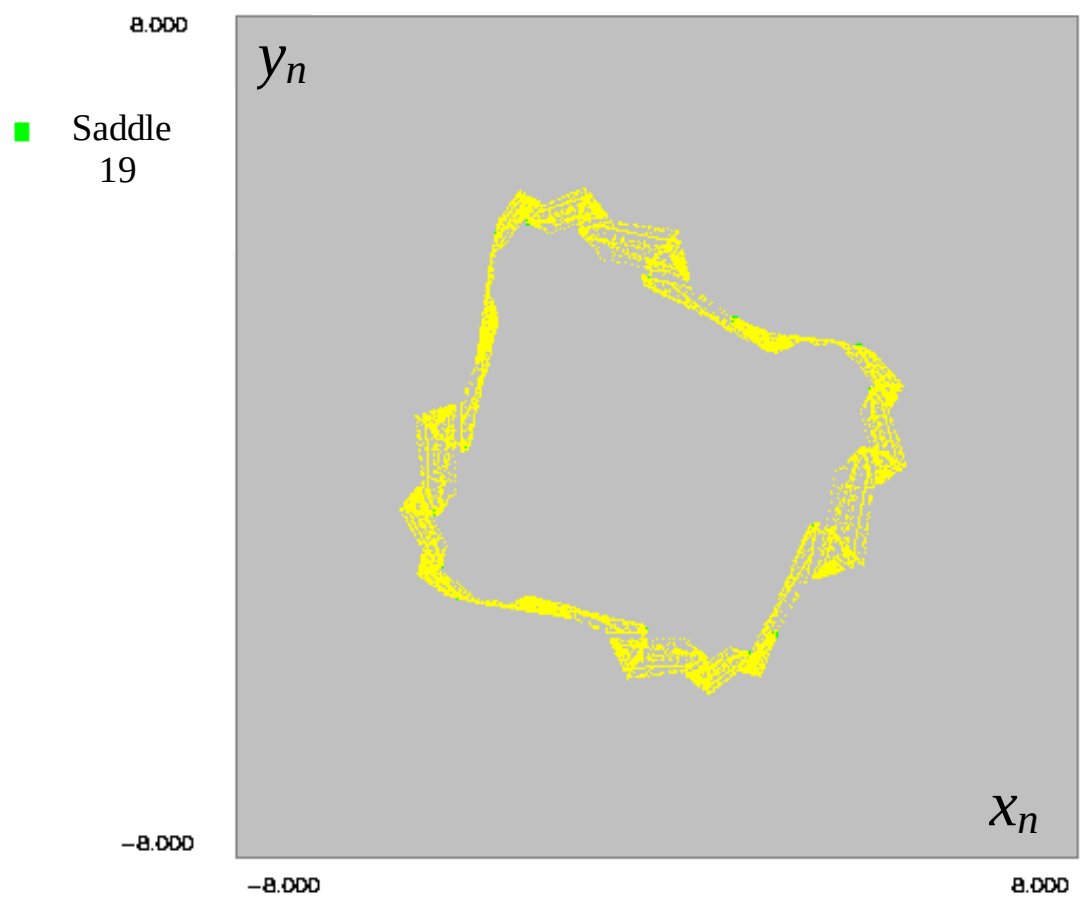

Fig. 19 : A strange repellor exists

$a_1$=-.22 $a_2$=-1.195 p=1.2 s=8

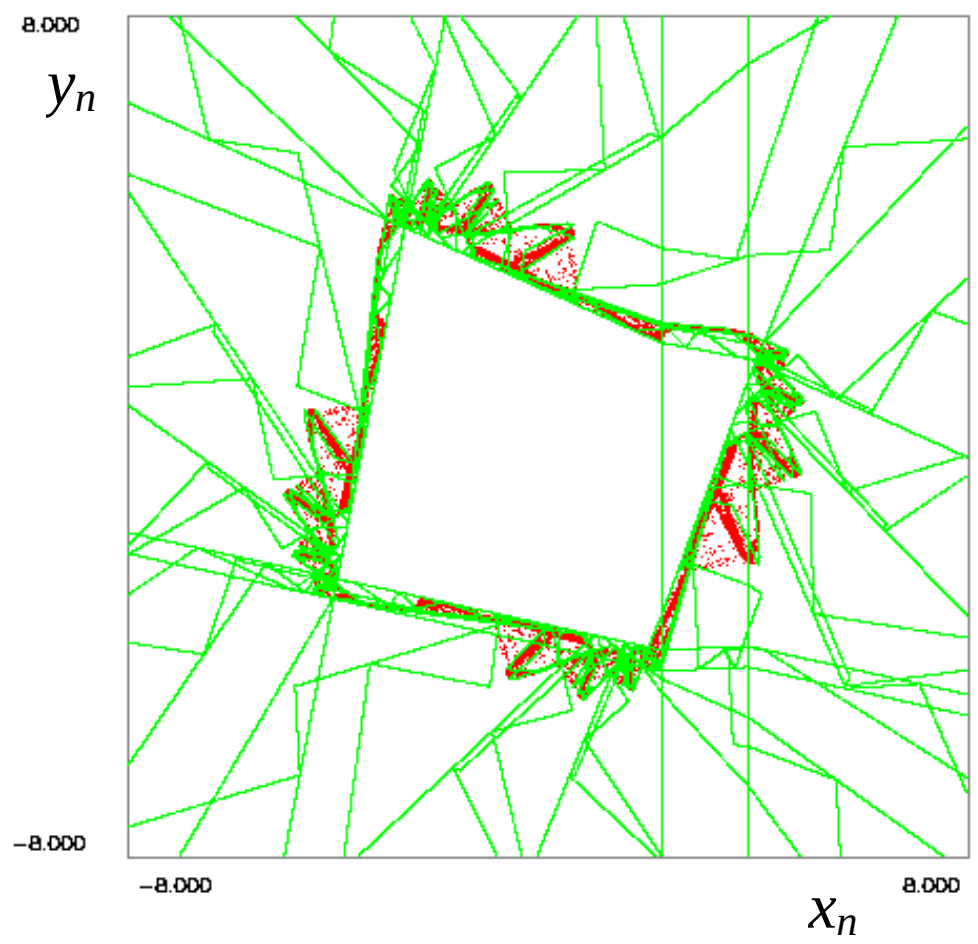

Fig. 20a : The chaotic attractor (in red) is delimited by segments of critical lines and their iterates.

$a_1$=-.2 $a_2$=-1.195 p=1.2 s=8

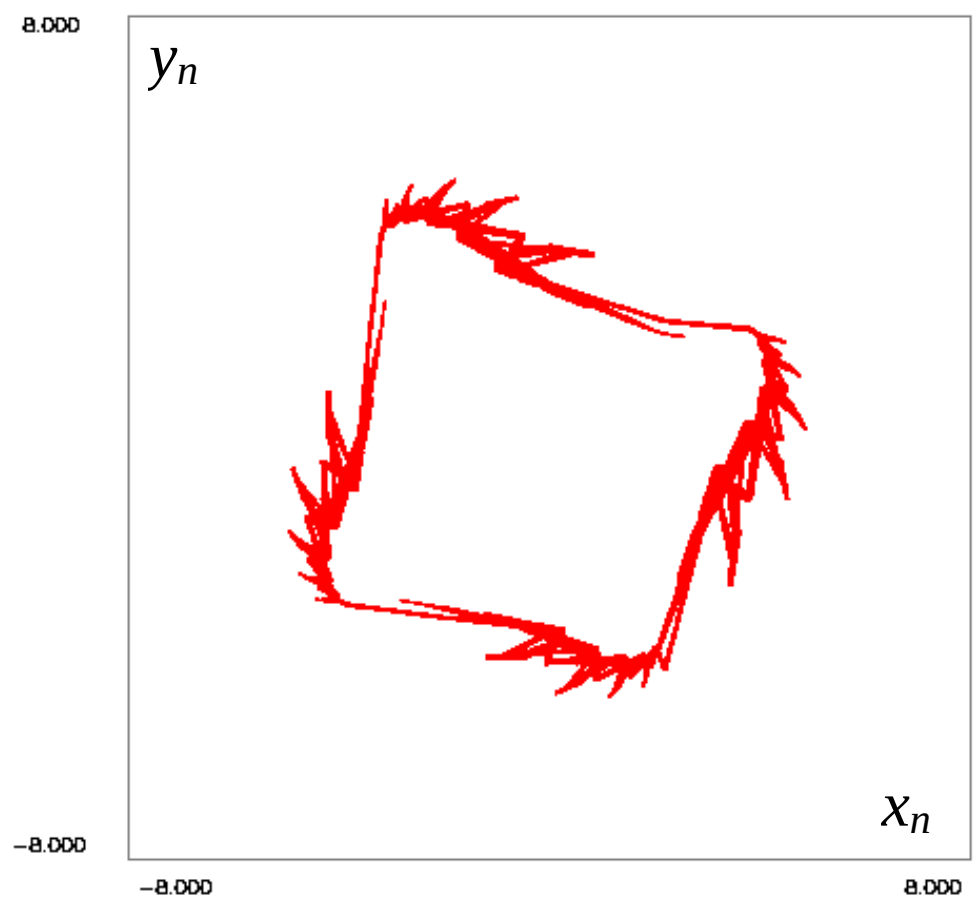

Fig. 20b : The strange repellor has become a chaotic attractor.

$a_1$=-.3221 $a_2$=-1.19486 p=1.2 s=8

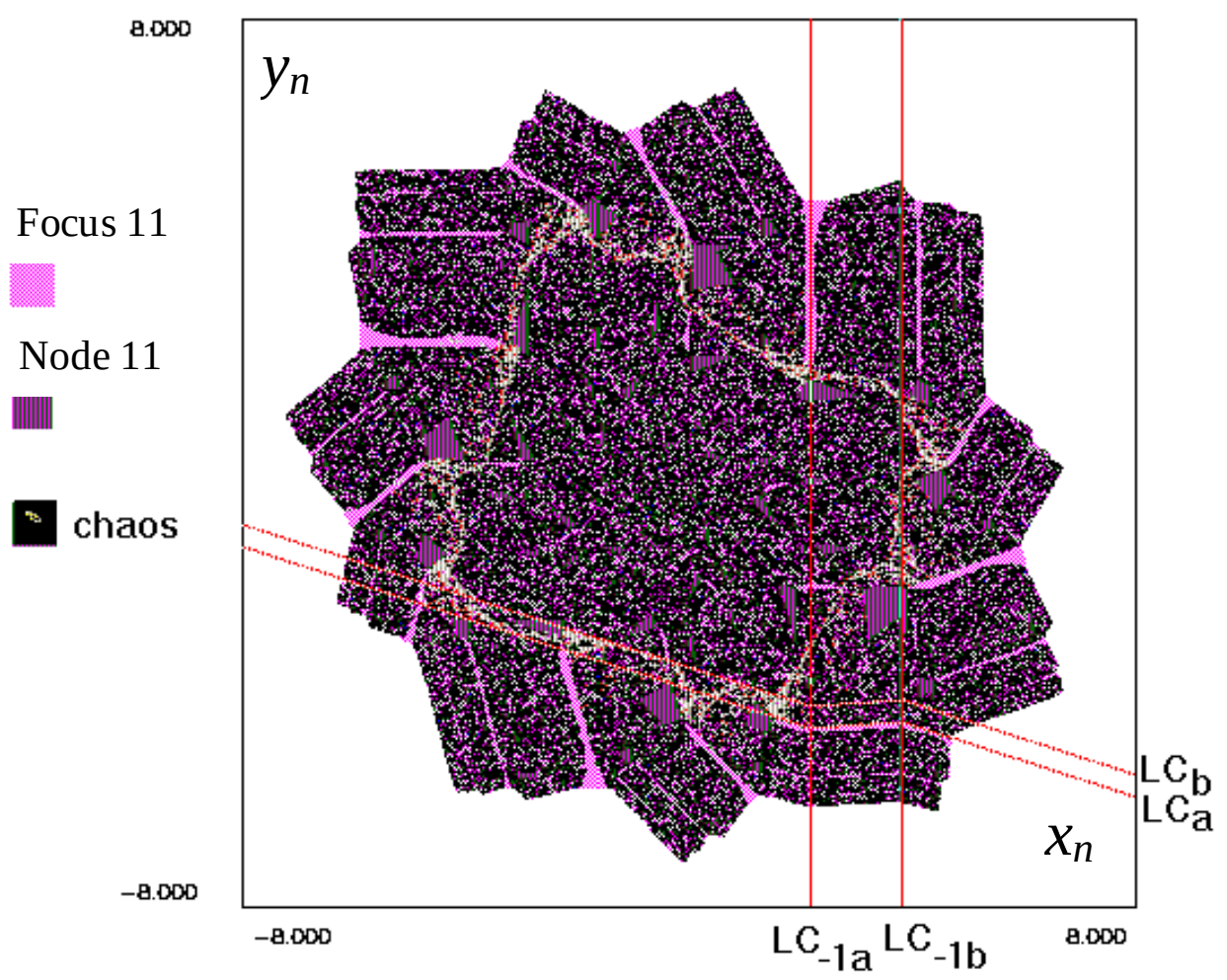

Fig. 21 Chaotic attractor coexists with two periodic orbits, the basin boundaries are fractal.

$a_1$=-.321 $a_2$=-1.19486 p=1.2 s=8

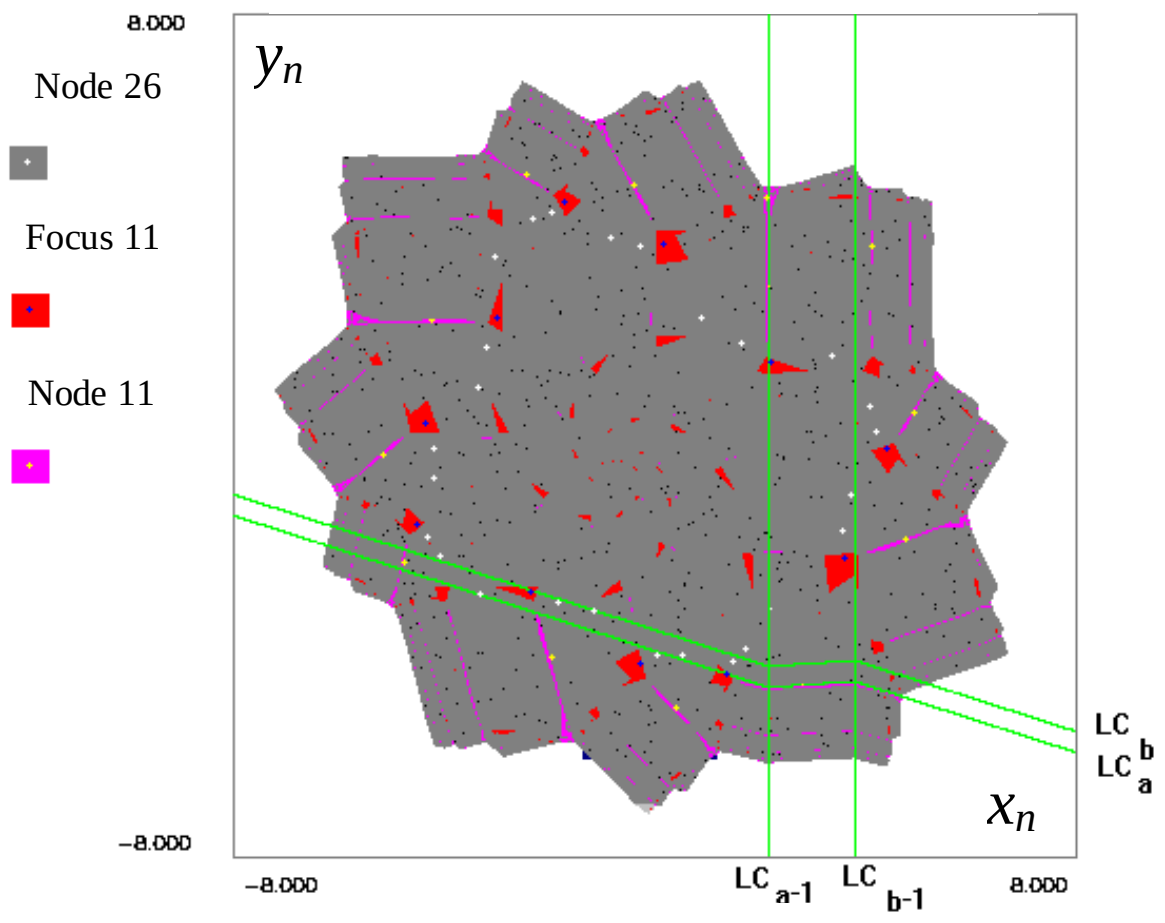

Fig.22 : Two period-11 orbits with the same rotation number : they belong to the same “tongue”.

$a_1$=.2482 $a_2$=-1.19486 p=1.2 s=8

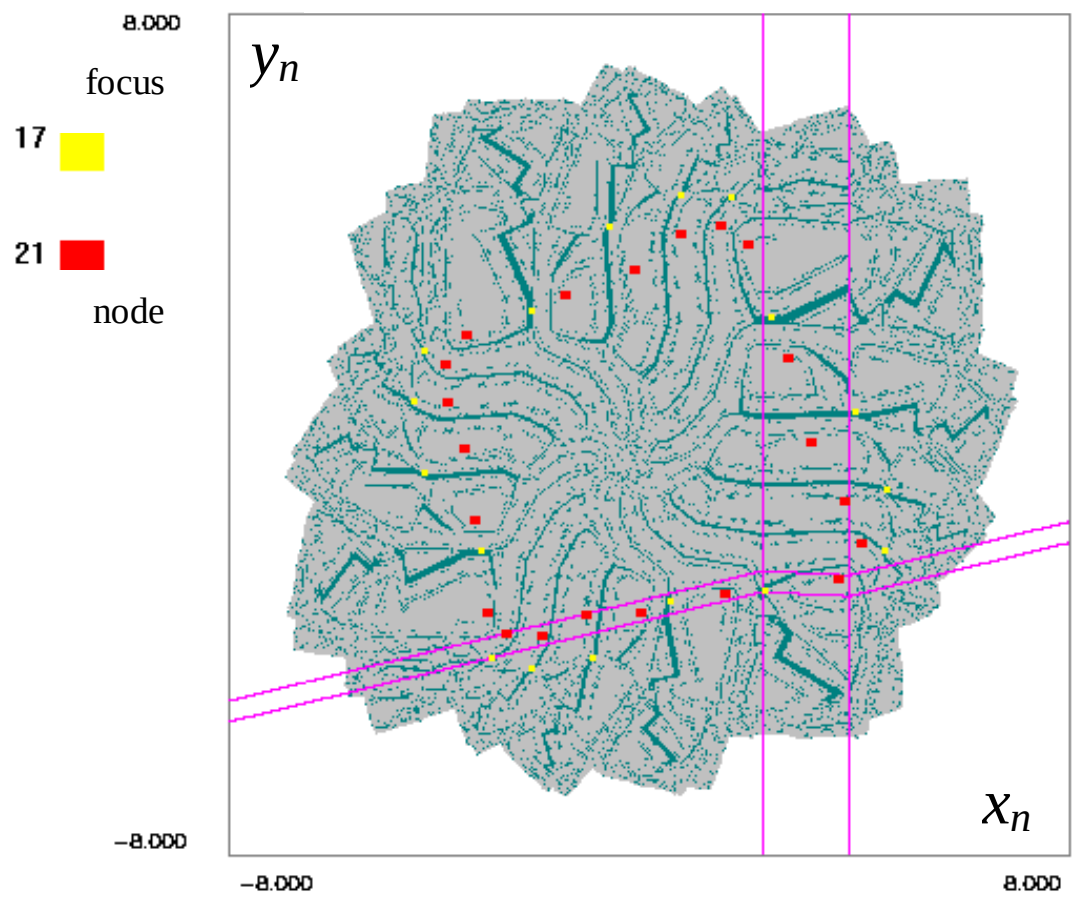

Fig. 23 : Fuzzy frontier for positive $a_1$.

In all cases of *fuzzy* frontier, the frontiers of the different basins bound regions which are intermingled one inside the other in a fractal structure, covering a whole region in the phase plane.

Then the problem of the initialization is extended to this region, as shown on Figs.16,17,21,22,23. Let us consider in details Fig. 17. The grey color corresponds to the initial conditions leading to divergent trajectories, the white to the basin of the period-15 focus and the region (in red) inside to the 3 intermingled basins of a period-15 node, a period-15 focus and a period-19 node. The set of these three basin boundaries is fuzzy. In the considered case it is difficult to predict whether the trajectories of the system will be captured by the period-19 node, the period-15 node or by the period-15 focus. This type of frontier means that the system evolution is unpredictable, and is caused by the finite precision achieved in the state vector initialization. It should be emphasized that, unlike the stochastic systems, as a result of the deterministic nature of our model in the ideal case, two identical starting conditions would generate identical trajectories.

Fig. 22 shows the basins of attraction of three different periodic orbits. The basins of both period-11 orbits (in red and magenta) are non-connected and form a set of *islands* inside the basin of the period-26 node (in dark grey) which is multiply connected (connected with *holes*). The period-11 focus and the period-11 node possess the same rotation number; a peculiarity of Arnold tongues structure can be brought out: each tongue is associated to a fixed rotation number [Boyland, 1986] [Mira, 1987], but each tongue can correspond to the existence of one or more attractive orbit (with the same period); in this case, two stable periodic orbits of period-11 coexist.

After decreasing $a_1$, a cascade of bifurcations occurs, giving rise to a chaotic attractor (Fig. 21). All basin frontiers become fuzzy. The phase portrait has changed drastically, although the global form of the basin has been kept. The period-26 node has disappeared, and the two period-11 orbits coexist with the chaotic attractor (whose basin is in black). The critical lines and their forward iterates delimit the latter.

It is worth noting the very small change in the parameter $a_1$ leading to the different phase portraits of Fig.21 and Fig.22. Despite the existence of stable areas for other fixed parameter vector values, the closeness of different tongues affects the stability of the encoder.

## 6. CONCLUSION

Our objective in this work has been to investigate the nonlinear dynamics of the DPCM system, resulting from the piece-wise modelling of the quantizer and to compare that model with the previously studied differentiable one. Through the parameter and phase plane study, different aspects of the dynamics of the DPCM system have been analyzed. Such a study is important to understand the real system, because bifurcations can be observed when parameters are varied and it is possible that parameter values change during the use of the system. Due to the very high parameter sensitivity shown in our paper, it is possible to observe chaotic behaviour, even when not expected. Nevertheless, the obtained results are robust in the sense that it can be expected that higher order transmission systems (with higher order predictor) display similar behaviour and that unpredictability and sensitivity with respect to initial conditions may be ubiquitous in the application of DPCM transmission techniques. Simulation and expectations from theoretical results support this observation. As seen at the system output, chaos with small amplitude (of the order of the noise in the system) can easily be mistaken for random noise. New phenomena arising from the chosen model have been summarized. The influence of changes of filter parameters on system dynamics has also been studied and the following results have been obtained :

1) As in the differentiable case :

- the dynamics of the system are extremely complex, and are characterized by extraordinary sensitivity to small changes of parameters,
- exact conditions for bifurcation instabilities and qualitative descriptions of the transient dynamics have been given,
- the parameter bifurcation diagram reveals a typical devil's staircase structure (or boxes in files),
- similar types of bifurcations occur, whatever the value of the constant input signal $s$; other studies have been done with a sinusoidal input signal which have shown analogous bifurcations [Rouabhi, Fournier-Prunaret, 1999],
- multistability has been brought out,
- stable chaotic behaviour can be observed,
- it is possible to tune the predictor parameters in such a way that oscillations of any chosen period could be generated (because of the boxes in files structure),

2) Unlike the differentiable case :

- classical bifurcations occur in a non traditional way,
- up to 4 coexisting attractors in our study can coexist, some of them having the same rotation number, (only three were obtained in the differentiable case),
- instead of differentiable Invariant Closed Curves, Weakly Chaotic Rings have been observed, due to the piece-wise linear structure of the model,
- the maximum amplitude of these chaotic oscillations can be predicted analytically.

Further studies should be devoted to analyze under what conditions robustness and even optimal performances can be guaranteed, despite the presence of nonlinearities and the lack of exact parameter and state vector initializations.

In fact it may be argued that it is exactly the nonlinear interaction between the signal estimation and the quantified error that give the whole system a measure of robustness. Further studies in this direction, comparing the DPCM and the Sigma-Delta modulator, under development.

**Acknowledgment**

We would like to thank the reviewers for their very useful and interesting comments and remarks. They have contributed significantly to improve the content of the paper. We also thank Dr.O. Feely for her helpful comments.